\documentclass{article}

\usepackage{amsmath}
\usepackage{amssymb}
\usepackage{tikz}
\usepackage{graphicx}
\usepackage[footnotesize]{caption}
\usepackage{subfig}
\usepackage{color}
\usepackage{rotating}
\usepackage{mathtools}
\usepackage{xcolor}
\usepackage[margin=0.85in]{geometry}

\usepackage{pifont}
\usetikzlibrary{decorations.pathmorphing}
\usetikzlibrary{arrows.meta}
\usetikzlibrary{calc}
\usetikzlibrary{shapes.misc}
\usepackage{subfig}

\usepackage[square,comma,numbers,sort&compress]{natbib}
\usepackage{hyperref}

\DeclareMathOperator{\erf}{erf}
\DeclareMathOperator{\Ai}{Ai}
\DeclareMathOperator{\Bi}{Bi}
\DeclareMathOperator{\Arg}{Arg}

\renewcommand{\Re}{\operatorname{Re}}
\renewcommand{\Im}{\operatorname{Im}}
\newcommand{\eps}{\epsilon}

\newcommand{\pdiff}[2]{\dfrac{\partial #1}{\partial #2}}
\newcommand{\diff}[2]{\dfrac{\mathrm{d} #1}{\mathrm{d} #2}}

\renewcommand{\i}{\mathrm{i}}
\newcommand{\e}{\mathrm{e}}

\usepackage{authblk}
\title{Appearance of the higher-order Stokes phenomenon in a discrete Airy equation}
\author[1]{Aaron J. Moston-Duggan\footnote{Corresponding Author. Electronic address: aaron.moston-duggan@hdr.mq.edu.au}}
\author[2]{Christopher J. Howls\footnote{Electronic address: c.j.howls@soton.ac.uk}}
\author[3]{Christopher J. Lustri\footnote{Electronic address: christopher.lustri@sydney.edu.au}}
\affil[1]{School of Mathematical and Physical Sciences, Macquarie University, Sydney, New South Wales, 2109, Australia}
\affil[2]{Mathematical Sciences, University of Southampton, Highfield, Southampton, SO17
1BJ, United Kingdom}
\affil[3]{School of Mathematics and Statistics, University of Sydney, Sydney, New South Wales, 2006, Australia}
\date{}

\begin{document}

\maketitle

\begin{abstract}
We study a discrete variant of the Airy equation, formulated as an advance-delay equation, to reveal that discretization induces the higher-order Stokes phenomenon, which is not present in the continuous Airy function and is typically only encountered in solutions to third-order or higher linear homogeneous, or nonlinear, differential equations. Using steepest descent and direct series methods, we derive asymptotic solutions and the Stokes structure. Our analysis shows that discretization produces a more intricate Stokes structure, containing higher-order Stokes phenomena and infinite accumulations of Stokes and anti-Stokes curves. The latter feature is a strictly nonlinear effect in continuous differential equations. We show that this unusual behavior can be generated in a discrete equation from a linear discretization. Numerical simulations confirm the  predictions, and a direct comparison with the continuous Airy equation explains how the discretization alters the Stokes structure.
\end{abstract}

\section{Introduction}

\begin{figure}[tb!]
    \centering
    \subfloat[$\epsilon=h=0.08$]{
        \includegraphics[width=0.48\textwidth]{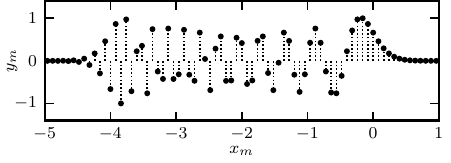}} 
    \subfloat[$\epsilon=h=0.04$]{
        \includegraphics[width=0.48\textwidth]{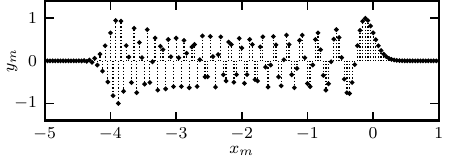}}
        
    \subfloat[$\epsilon=h=0.02$]{
        \includegraphics[width=0.48\textwidth]{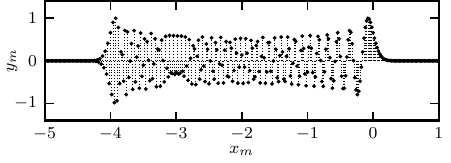}} 
    \subfloat[$\epsilon=h=0.008$]{
        \includegraphics[width=0.48\textwidth]{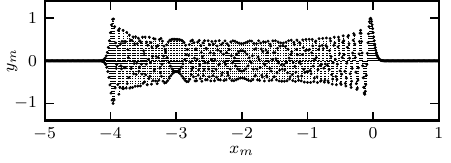}}
    \caption{Numerical solutions $y_m$ of the discrete Airy equation \eqref{eqn:latAi}, which decay as $x_m \to \pm \infty$, for several values of $\epsilon$ and $h$. Each solution is computed with $y_0 = 1$ at $x_0 = -2$ and rescaled so that the maximum value is $1$ for comparison. The solutions are oscillatory with slowly varying amplitude between the turning points at $x = 0$ and $x = 4$, and decay exponentially outside this region.}
    \label{fig:dAiNSol}
\end{figure}

The Stokes phenomenon \cite{stokes1864} describes the sudden appearance of exponentially small contributions across Stokes curves in the complex plane. A more recent discovery, the higher-order Stokes phenomenon \cite{chapman2005,howls2004}, describes the appearance or termination of such curves; in continuous settings it appears only in homogeneous linear differential equations of third-order or higher. We will see that simply discretizing a homogeneous second-order differential equation can generate the higher-order Stokes phenomenon.  Furthermore, the solution will contain Stokes curve accumulations, which are never seen in homogeneous linear differential equations of finite order and are usually regarded as a nonlinear effect. Our results reveal a striking fact: a linear discretization of a linear differential equation can produce phenomena usually associated with higher-order or nonlinear continuous systems.

The Airy function that decays exponentially as $x\rightarrow +\infty$ is the original, and canonical, example of the Stokes phenomenon \cite{stokes1864} and has a well-known Stokes curve structure. A scaled version of the function satisfies the singularly perturbed continuous differential equation, 
\begin{equation}
    \label{eqn:spAi}
        \epsilon^2\diff{^2y}{x^2}-xy=0 \quad\quad \text{as} \quad\quad \epsilon\to0 \, ,
\end{equation}
 A discretization of this equation may be obtained by applying a second-order central difference to the derivative terms in \eqref{eqn:spAi}, to obtain
\begin{equation}
\label{eqn:latAi}
\dfrac{\epsilon^2}{h^2}\left(y_{m+1} - 2y_m + y_{m-1}\right) - x_m y_m = 0 \quad \text{as} \quad \epsilon \to 0 \,,
\end{equation}
where $0 < \epsilon \ll 1$, and $m \in \mathbb{Z}$ indexes solution values $y_m$ defined at points $x_m$. The lattice spacing $h = x_{m+1} - x_m$, may vary for fixed $\epsilon$ and is $\mathcal{O}(\epsilon)$ in our analysis. We henceforth refer to \eqref{eqn:latAi} as the ``discrete Airy equation''.

Figure \ref{fig:dAiNSol} shows numerical solutions to \eqref{eqn:latAi} for $y_m \to 0$ as $|x_m| \to \pm \infty$, computed using the method from Section \ref{sect:NUM}. These results motivate the present study, revealing a central oscillatory region with slowly varying amplitude bounded by exponentially decaying outer regions. This already contrasts with the solutions of the continuous Airy equation \eqref{eqn:spAi}, which feature only one oscillatory region and one (growing or decaying) exponential region. 

We will later derive the asymptotic solution to \eqref{eqn:latAi} and find two turning points at $x = -4$ and $x = 0$, in contrast to the one turning point at $x=0$ in the continuous Airy equation \eqref{eqn:spAi}. Stokes switching across these points produces the central oscillatory region shown in Figure \ref{fig:dAiNSol}, and the asymptotic results match the numerical solutions.

By analysing the Stokes structure of solutions to \eqref{eqn:latAi}, we show that they exhibit the higher-order Stokes phenomenon and infinitely many Stokes and anti-Stokes curves accumulating onto limiting curves in the complex plane. Such features are typically absent in linear homogeneous second-order ordinary differential equations, instead arising in inhomogeneous \cite{shelton2023}, nonlinear \cite{chapman2007,honda2007,honda2008}, higher-order \cite{aoki2005, honda2007, honda2008}, or partial \cite{chapman2007,howls2004} differential equations, and, to our knowledge, have not previously been observed in difference equations. We demonstrate that this behavior arises from discretization and can therefore occur even in homogeneous linear second-order discrete equations. In contrast, the continuous Airy equation \eqref{eqn:spAi}, as a homogeneous linear second-order ordinary differential equation, does not exhibit these features.

We shall study the transition between the fully discrete and original equation via a continuum approximation to the discrete Airy equation \eqref{eqn:latAi} with $x=hm$ and $y(x)=y_m$ to obtain
\begin{equation}
    \label{eqn:soadvAi}
        \dfrac{1}{\sigma^2}\left(y(x+\sigma\epsilon)-2y(x)+y(x-\sigma\epsilon)\right)-   xy(x)=0 \quad\quad \text{as} \quad\quad \epsilon\to 0 \, ,
\end{equation}
where $h = \sigma\eps$. We refer to \eqref{eqn:soadvAi} as the advance-delay Airy equation, and we determine its asymptotic solutions using the steepest descent method \cite{bender2013}.

The method of steepest descent  \cite{bender2013}  is well suited to linear problems, however, unless recast in terms of a Borel transform, it is generally not applicable to nonlinear discrete equations.  Instead such problems may be approached using factorial-over-power methods \cite{dingle1973,chapman1998}. We repeat the analysis of \eqref{eqn:soadvAi} in Appendix \ref{sect:EA} using factorial-over-power methods, and show that it produces identical asymptotic solutions and Stokes structure.

The discrete Airy equation \eqref{eqn:latAi} arises in implementing transparent boundary conditions \cite{arnold1998b, arnold1998a, ehrhardt2004,tappert1977} and non-standard discretization schemes \cite{mickens1993, mickens1997} used in numerical methods for wave propagation problems, including acoustics \cite{arnold1998b, ehrhardt2004} and electromagnetism \cite{arnold1998a}. The studies \cite{ehrhardt2004,mickens1997} derive asymptotic solutions to \eqref{eqn:latAi} as $m\to \infty$ for $\epsilon=1$ and $x_m = hm$, while \cite{barnes1904,ehrhardt2004} obtain additional solutions. The authors of \cite{ehrhardt2004} discuss inconsistencies in the asymptotic results in \cite{barnes1904,mickens1997, ehrhardt2004}.

The studies \cite{barnes1904,ehrhardt2004,mickens1997} use classical asymptotic methods, which apply to the discrete Airy equation \eqref{eqn:latAi} only for specific restrictions on variables and parameters. The authors in \cite{wimp1985,wong2022} discuss these methods, and note that they may be incomplete. Here, we extend the asymptotic analysis of \eqref{eqn:latAi} into the complex plane without any restrictions. This reveals the significance of the previously unidentified higher-order Stokes phenomenon and demonstrates the accumulation of curves in the Stokes structure.

This paper proceeds as follows. In \ref{sect:BD} we recall the concept of the Stokes and higher order Stokes phenomenon and previous work on the asymptotics of discrete equations.  In Section \ref{sect:SD}, using the steepest descent method, we derive the asymptotic solutions and the Stokes structure of the advance–delay Airy equation \eqref{eqn:soadvAi}. Section \ref{sect:NUM} presents numerical solutions to the discrete Airy equation \eqref{eqn:latAi} and compares them with the asymptotic results of \eqref{eqn:soadvAi} from Section \ref{sect:SD} and the literature. In Section \ref{sect:CON}, we discuss our results and conclude. Appendix \ref{sect:EA} reproduces the results of Section \ref{sect:SD} using the factorial-over-power method, applicable to both linear and nonlinear difference equations.

\section{Definitions and Previous Work}\label{sect:BD}

\subsection{The Stokes and higher-order Stokes phenomenon }

A Stokes phenomenon may occur between pairs of exponentially prefactored asymptotic contributions with distinct exponents $\chi_1$ and $\chi_2$, 
\begin{equation}\label{eqn:sl_cond}
\mathrm{Im}\left(\chi_1 - \chi_2\right) = 0 \, ,
\end{equation}
If $\mathrm{Re}(\chi_1)>\mathrm{Re}(\chi_2)$ as the Stokes line is crossed, the maximally dominant asymptotic contribution involving $\chi_1$ may switch on a subdominant contribution involving $\chi_2$.  Anti-Stokes lines occur where $\mathrm{Re}\left(\chi_1 - \chi_2\right) = 0$, at which neither contribution dominates.

A classic example of the Stokes phenomenon occurs in the solution of the continuous Airy equation \eqref{eqn:spAi} in terms of the eponymous functions $\Ai$ and $\Bi$ \cite{NIST:DLMF}. The general solution of \eqref{eqn:spAi} is
\begin{equation}\label{eq:fullairy}
    y = C_1 \Ai(\epsilon^{-2/3}x) + C_2 \Bi(\epsilon^{-2/3}x) \, ,
\end{equation}
where $C_1$ and $C_2$ are arbitrary constants and $\Ai(\epsilon^{-2/3}x)$ and $\Bi(\epsilon^{-2/3}x)$ are linearly independent solutions.

The Stokes structure of \eqref{eqn:spAi} can be obtained, for example, by applying the steepest descent method to an integral representation of the solution. The asymptotic solution has two saddle points at $z_1 = -\mathrm{i}x^{1/2}$ and $z_2 = \mathrm{i}x^{1/2}$, giving the asymptotic contributions
\begin{equation}
    \label{eqn:spAiasy11}
        y_1 \sim  \frac{-\i C_3}{2\sqrt{\pi\eps}x^{1/4}}\mathrm{e}^{2x^{3/2}/3}\quad \mathrm{and} \quad  y_2 \sim  \frac{ C_3}{2\sqrt{\pi\eps}x^{1/4}}\mathrm{e}^{-2x^{3/2}/3} \quad \text{as} \quad \epsilon \to 0 \, .
\end{equation}
These contributions match the exact solution \eqref{eq:fullairy} for $C_2 = 0$ and $C_3 = C_1\epsilon^{2/3}$, after applying the scaling for \eqref{eqn:spAi}.

\begin{figure}[tb!]
    \centering
    \includegraphics[width=0.75\linewidth]{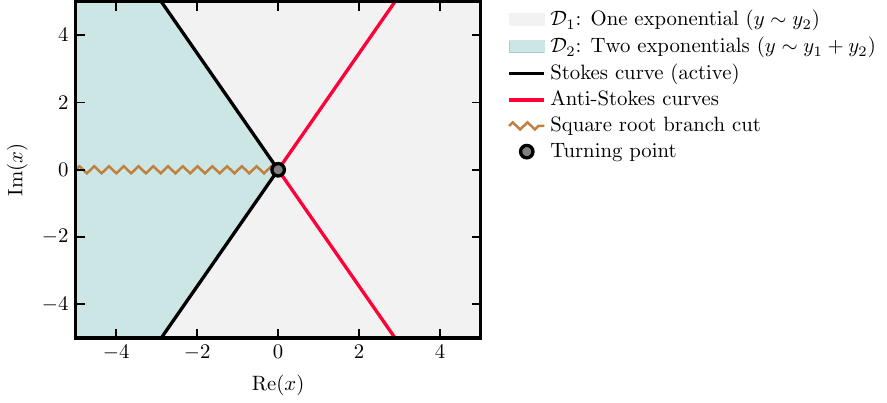}
    \caption{Stokes structure of the $\Ai$ solution to the Airy equation \eqref{eqn:spAi}. Stokes (black) and anti-Stokes (red) curves emerge from the turning point at $x = 0$. In region $\mathcal{D}_1$, the solution has a single decaying exponential as $x \to \infty$ on the real axis. Crossing the anti-Stokes curve causes this contribution to grow instead of decay. Crossing the Stokes curve into $\mathcal{D}_2$ causes a second exponential contribution to appear, resulting in oscillatory behavior on the negative real axis.}
    \label{fig:Aistokstruc}
\end{figure}

We may observe from \eqref{eqn:spAiasy11} that $\chi_1=2/3x^{3/2}$, $\chi_2=-2/3x^{3/2}$ and use \eqref{eqn:sl_cond} to derive Figure \ref{fig:Aistokstruc}, which shows a schematic of the Stokes structure containing the Stokes curves, the anti-Stokes curves, and a description of the asymptotic contributions from $y_1$ and $y_2$ \eqref{eqn:spAiasy11} in each region of the complex $x-$plane. For the asymptotic approximation to the specific solution $\Ai(\epsilon^{-2/3}x)$, the Stokes curves with $\Arg(x) = \pm 2\pi/3$ are active, while there is no Stokes curve on the positive real axis (since there is no exponentially growing/dominant expansion for ${\rm Re} \ x>0$). As the active Stokes curves are crossed from right to left, the $y_2$ contribution switches on the $y_1$ contribution.

In this study, we consider the generalisation of the Stokes phenomenon known as the higher-order Stokes phenomenon \cite{chapman2005, howls2004, shelton2023, Takei2017}, the effects of which were first noted in \cite{berk1982}. The higher-order Stokes phenomenon can occur when a solution contains three distinct exponential contributions with exponents $\chi_1$, $\chi_2$, and $\chi_3$. As shown in \cite{chapman2005,howls2004}, it occurs when the exponents satisfy the condition
\begin{equation}\label{eqn:hosl_cond}
\mathrm{Im}\left(\frac{\chi_1 - \chi_2}{\chi_1 - \chi_3}\right) = 0 \, ,
\end{equation}
which corresponds to colinearity of singularities in the Borel plane \cite{howls2004}. The higher order Stokes curves that satisfy \eqref{eqn:hosl_cond}, are generated by Stokes crossing points, where three Stokes curves intersect. Crossing such higher order curves alters which saddle points are adjacent, thereby affecting the strength of the Stokes switching, even leading to the truncation of the ordinary Stokes curves at the Stokes crossing (and regular) point.

The higher-order Stokes phenomenon has been studied using both WKB methods and direct series methods, with reviews given in \cite{aoki2001, honda2015, Takei2017} for WKB approaches and \cite{shelton2023} for series methods. 

Clearly a higher order Stokes phenomenon cannot occur for the continuous Airy equation \eqref{eqn:spAi} or any homogenous linear second-order ordinary differential equation.  However, here we show that the higher-order Stokes phenomenon is a key feature of the asymptotic solutions to the discrete Airy equation \eqref{eqn:soadvAi}.

We use condition \eqref{eqn:hosl_cond} to identify higher-order Stokes curves in the asymptotic solutions of \eqref{eqn:soadvAi} and examine their impact on the solutions behaviour and how this change of adjacency impacts the steepest descent analysis.

\subsection{Asymptotics of discrete equations}

The discrete Airy equation \eqref{eqn:latAi} belongs to the class of second-order difference equations, also known as three-term recurrence relations. Such equations appear in special function theory \cite{joshi2015, wang2003}, orthogonal polynomials \cite{huang2020,li2020, wang2012}, random matrix theory \cite{dai2014}, quantum mechanics \cite{braun1978}, lattice dynamics \cite{alfimov2019, burr2000, lustri2020, moston2022}, and topological string theory \cite{kashani2016}. These applications have motivated the development of asymptotic methods for discrete systems, broadly classified into two approaches: direct series methods and WKB methods.

The series methods developed in \cite{cao2014,li2020,wang2003,wang2005,wong1992b,wong2014,wong2022} provide asymptotic solutions to second-order difference equations with linear potentials and coefficients in $x_m$ in the limit $|m|\to\infty$. These approaches extend classical asymptotic methods: see \cite{wong2014, wong2022} for reviews. The systems considered in these studies contain two special points in $x_m$, termed turning points by analogy with WKB theory \cite{bender2013}. In each of these problems it is possible to construct solutions which are oscillatory with a slowly varying envelope between the turning points, while outside this region they grow or decay exponentially \cite{cao2014, li2020, wang2003, wang2005}. 

WKB methods for second-order difference equations have been developed using discrete and continuous approaches. Studies such as \cite{spigler1994, spigler2019} provide a background of these approaches and propose a unified framework bridging the two. These results are rigorously developed in \cite{geronimo1992, geronimo2004}, where asymptotic error bounds are derived. As in the series solutions, the WKB solutions predict two turning points for linear second-order difference equations. 

Discrete versions of more advanced WKB methods, such as complex and exact WKB, have been explored in, for example, \cite{burr2000, fedotov2019,kashani2016} and allow asymptotic analysis of discrete equations with complex domains. Extending the domain into the complex plane enables the study of the Stokes phenomenon present in the solution. These prior studies determine some aspects of the Stokes phenomenon in second-order difference equations. Our results will build on these ideas by providing a complete picture of the Stokes structure, including the higher-order Stokes phenomenon. 

\section{Steepest Descent Analysis of the Discrete Airy Equation}
\label{sect:SD}

In this section, we compute asymptotic solutions of the advance-delay Airy equation \eqref{eqn:soadvAi} using the steepest descent method from \cite{bender2013}, and determine the associated Stokes structure.   

The steepest descent method used here is typically only possible for linear (or linearizable) equations. More general exponential asymptotic methods have been developed for nonlinear equations (e.g. \cite{chapman1998,howls2004,king2001,olde1995}). Appendix \ref{sect:EA} illustrates how the factorial-over-power method from \cite{chapman1998, king2001,olde1995} can reproduce the results from this section in a fashion that could be applied directly to nonlinear difference equations.

\subsection{The steepest descent method}

We apply the Fourier transform \eqref{eqn:foutfm} 
\begin{equation}
    \label{eqn:foutfm}
        \hat{y}(\omega) = \int_{-\infty}^\infty y(x) \mathrm{e}^{- \mathrm{i} \omega x} \, \mathrm{d}x \quad \text{and}            \quad y(x) = \dfrac{1}{2\pi} \int_{-\infty}^{\infty} \hat{y}(\omega) \mathrm{e}^{\mathrm{i} \omega x} \, \mathrm{d} \omega \, .
\end{equation}
to the advance-delay Airy equation \eqref{eqn:soadvAi} and solve the resulting equations for $\hat{y}$, setting $z = \sigma\epsilon\omega$ and using the inverse transform gives a representation of the exact solution $y$ in the form
\begin{equation}
\label{eqn:SDAns}
    y = \int_{-\infty}^\infty g(x,z)\mathrm{e}^{\phi(x,z)/\epsilon}        \,\mathrm{d}z \quad\quad \text{as} \quad\quad \epsilon \to 0 \, ,
\end{equation}
where
\begin{equation}\label{eqn:soadAiSDE}
 g(z) = \frac{C}{2\pi\sigma\epsilon} \quad \text{and} \quad \phi(x,z) = \frac{\mathrm{i}}{\sigma}\left(zx + \frac{2}{\sigma^2}(z+\mathrm{i}\sinh(\mathrm{i}z))\right) \, . 
\end{equation}
Here $C$ is an arbitrary constant. The exact solution \eqref{eqn:SDAns} is a Laplace-type integral, to which we can apply the steepest descent method to determine asymptotic solutions as $\epsilon\to 0$.

\subsubsection{Saddle point locations}

\begin{figure}[tb!]
    \centering
    \includegraphics[width=0.9\linewidth]{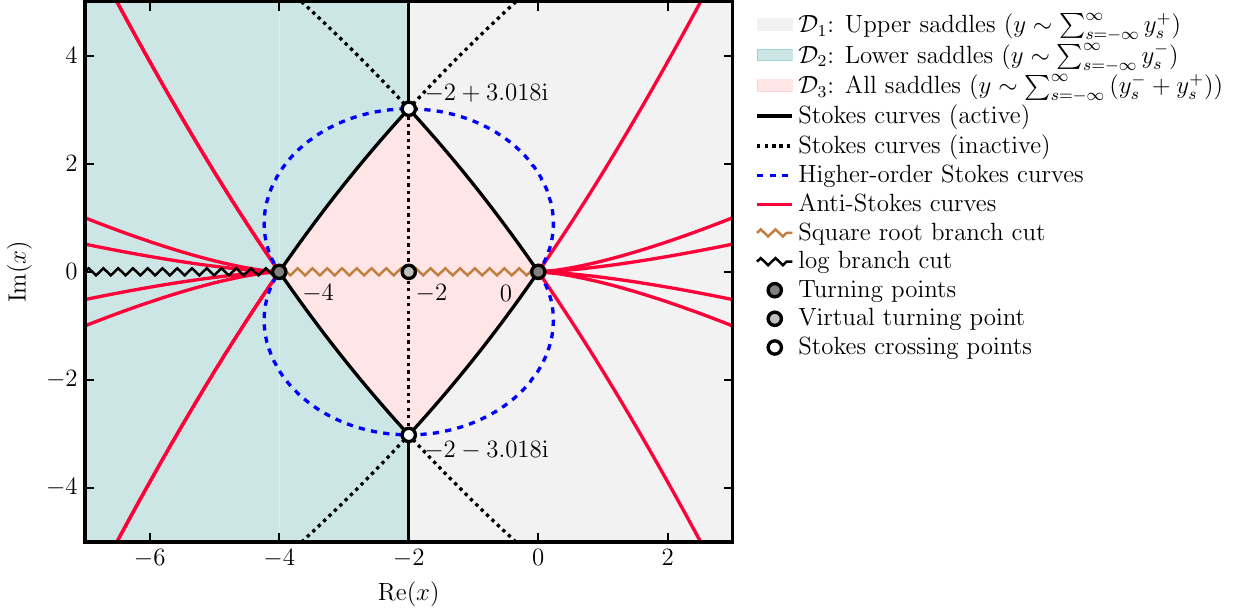}
    \caption{Stokes structure of the advance-delay Airy equation \eqref{eqn:soadvAi} for $\sigma=1$, with turning points at $x =  0$ and $x = -4$ (gray circles). In region $\mathcal{D}_1$, contributions from upper saddle points $z_s^+$ are exponentially decaying on the positive real axis but become growing after crossing anti-Stokes curves (red). An infinite number of anti-Stokes curves accumulate near the real axis, each for a different $z_s^+$ (only first three shown). Similar behaviour occurs for lower saddle points $z_s^-$ along the negative real axis. Crossing Stokes curves (black) causes contributions to switch on and off. In $\mathcal{D}_2$, contributions from $z_s^+$ vanish and $z_s^-$ appear. In $\mathcal{D}_3$, all contributions appear. Higher-order Stokes curves (dashed blue) truncate the active Stokes curves at the Stokes crossing points (white circles).}
    \label{fig:dAistokstruc}
\end{figure}

The saddle points of \eqref{eqn:SDAns} satisfy 
\begin{equation}
    \label{eqn:sadptcond}
   \pdiff{\phi(x,z_s)}{z} = 0  \quad \text{and} \quad \frac{\partial^2{\phi(x,z_s)}}{\partial z^2} \ne 0 \,,
\end{equation}
which yields from \eqref{eqn:soadAiSDE} an infinite number of simple saddle points located at
\begin{equation}
\label{eqn:singty}
z_s^- = -\mathrm{i}\cosh^{-1}\left(1+\frac{\sigma^2 x}{2}\right) + 2\pi s \quad \text{and} \quad z_s^+ =  \mathrm{i}\cosh^{-1}\left(1+\frac{\sigma^2 x}{2}\right) + 2\pi s \, ,
\end{equation}
where $s\in\mathbb{Z}$. These saddle points form two sets; $z_s^+$ in the upper-half $z$-plane and $z_s^-$ in the lower-half $z$-plane.

Finally, we must specify the branch structure of \eqref{eqn:singty}. The inverse hyperbolic cosine can be expressed as 
\begin{equation}
\cosh^{-1}\left(1+\frac{\sigma^2x}{2}\right) = \log\left(1+\frac{\sigma^2x}{2}+\frac{1}{2}\sqrt{\sigma^2 x}\sqrt{4+\sigma^2 x}\right) \, ,
\end{equation}
which has branch points at $x=-4/\sigma^2$, and $x=0$. We place the square root branch cut along the real interval $x \in (-4/\sigma^2, 0)$; crossing this cut changes the sign of one square root, mapping $z_s^+ \mapsto z_s^-$ and vice versa. The logarithmic branch cut lies along $x \in (-\infty, -4/\sigma^2)$; crossing thus cut from above adds  $2\pi\mathrm{i}$ to the logarithm, which shifts the saddle point index, such that $z_s^+ \mapsto z_{s-1}^+$ and $z_s^- \mapsto z_{s+1}^-$. 

\subsubsection{Saddle point contributions}

Each saddle point $z_s^\pm$ has an associated saddle height that we denote as $\phi_s^\pm$, given by
\begin{equation}\label{e:sheights}
    \phi_s^{\pm} (x) \coloneqq \phi(x,z_s^{\pm}) = \frac{\mathrm{i}}{\sigma }\left[\left(x+\frac{2}{\sigma^2}\right)\left(\pm\mathrm{i}\cosh^{-1}\left(1+\frac{\sigma^2x}{2}\right) + 2\pi s \right)\mp\mathrm{i}\sqrt{x\left({\sigma^2}x+4\right)}\right] \, , 
\end{equation}
where the upper and lower signs correspond. Throughout the remainder of this section, we use the same subscript and superscript notation to denote quantities associated with the saddle points $z_s^-$ and $z_s^+$ as defined in \eqref{eqn:singty}. 

The asymptotic contributions as $\eps \to 0$ are found using the saddle point formula from \cite{bender2013}, giving
\begin{align}
\label{eqn:dasy1}
y_s^{+} &\sim \frac{1}{\sqrt{2\pi\epsilon}x^{1/4}(\sigma^2x+4)^{1/4}}\mathrm{e}^{\left(\frac{\mathrm{i}}{\sigma \epsilon}\left[\left(x+\frac{2}{\sigma^2}\right)\left(\mathrm{i}\cosh^{-1}\left(1+\frac{\sigma^2x}{2}\right)  + 2\pi s \right) - \mathrm{i}\sqrt{x\left({\sigma^2}x+4\right)}\right]   \right)} \, , \\
\label{eqn:dasy2}
y_s^{-} &\sim \frac{\mathrm{i}}{\sqrt{2\pi\epsilon}x^{1/4}(\sigma^2x+4)^{1/4}}\mathrm{e}^{\left(\frac{\mathrm{i}}{\sigma \epsilon}\left[\left(x+\frac{2}{\sigma^2}\right)\left(-\mathrm{i}\cosh^{-1}\left(1+\frac{\sigma^2x}{2}\right)  + 2\pi s \right) + \mathrm{i}\sqrt{x\left({\sigma^2}x+4\right)}\right]   \right)} \, .
\end{align}
Each contribution $y_s^\pm$ has two singularities, hence, the solution has turning points at $x= -4/\sigma^2$ and $x = 0$. 

\subsubsection{Stokes, anti-Stokes and higher-order Stokes curves}

We introduce notation for (possibly inactive) Stokes curves $\mathcal{S}$ and anti-Stokes curves $\mathcal{A}$ due to interactions between contributions from saddle points $z_s^\pm$. All such pairings can be written as
\begin{align}
\mathcal{S}_{s,p}^{\pm,\pm} = \{x \in \mathbb{C} \,| \Im(\phi_s^{\pm} -\phi_p^{\pm}) = 0\} \quad \text{and} \quad  \mathcal{A}_{s,p}^{\pm,\pm} = \{x \in \mathbb{C}\,| \Re(\phi_s^{\pm}-\phi_p^{\pm}) = 0 \} \, ,
\end{align}
where the sign choices on the same side of the equality are independent, but the first sign choice on each side and the second sign choice on each side correspond. 

Using condition \eqref{eqn:hosl_cond}, we introduce notation for relevant higher-order Stokes curves, which we denote by
\begin{align}
    \mathcal{H}_{s,s+1,s+1}^{-,+,-} &= \left\{x \in \mathbb{C} \,\left|\, \Im\left(\frac{\phi_s^--\phi_{s+1}^+}{\phi_s^--\phi_{s+1}^-}\right) = 0 \right. \right\},\\  
    \mathcal{H}_{s,s,s+1}^{+,-,+} &= \left\{x \in \mathbb{C} \,\left|\, \Im\left(\frac{\phi_s^+-\phi_{s+1}^-}{\phi_s^+-\phi_{s+1}^+}\right) = 0 \right. \right\} \, .
\end{align}
While other triplets exist, only these higher-order Stokes curves influence the solution.

\subsection{Detailed analysis for \texorpdfstring{$\sigma=1$}{sigma = 1}} \label{stepdescsigone}

\begin{figure}[tb!]
    \centering
\subfloat[A typical steepest descent curve for $x \in \mathcal{D}_1$]{
        \includegraphics[width=0.48\textwidth]{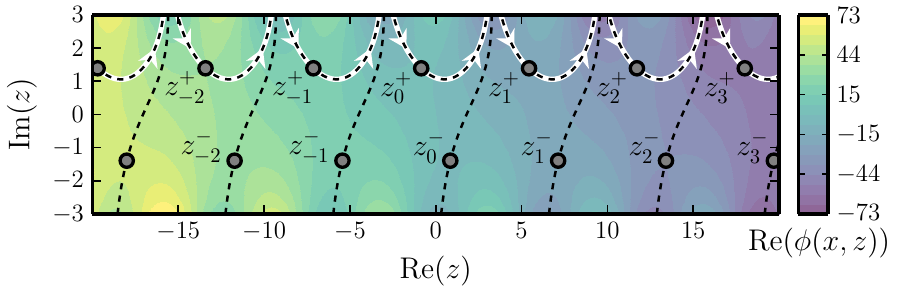}
        \label{fig:sdregd1}} 
\subfloat[A typical steepest descent curve for $x \in \mathcal{D}_2$]{
        \includegraphics[width=0.48\textwidth]{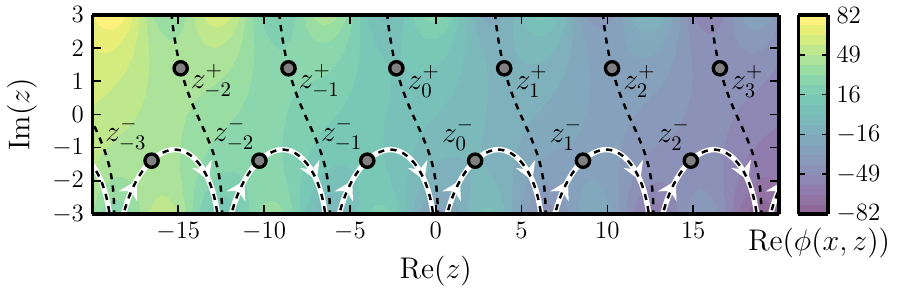}
        \label{fig:sdregd2}}
    
\subfloat[A typical steepest descent curve for $x \in \mathcal{D}_3$]{
        \includegraphics[width=0.48\textwidth]{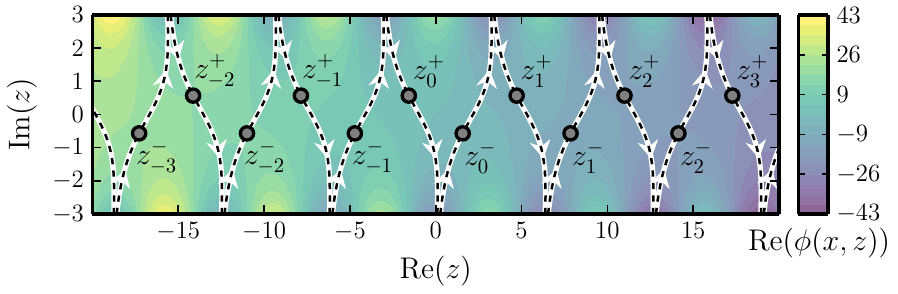}
        \label{fig:sdregd3}}
    \caption{Steepest descent analysis schematics for the advance-delay Airy equation \eqref{eqn:soadvAi} for $\sigma=1$, showing the integration path (white), saddle points $z_s^\pm$ (circles), and constant phase contours (dashed). Figures (a), (b), and (c) show typical schematics in regions $\mathcal{D}_1$, $\mathcal{D}_2$, and $\mathcal{D}_3$, respectively.}
    \label{fig:sdreg}
\end{figure}

In this section, we determine the Stokes structure and construct the asymptotic solution of the advance-delay Airy equation \eqref{eqn:soadvAi} for the case $\sigma = 1$. A schematic of the resulting Stokes structure is shown in Figure \ref{fig:dAistokstruc}.

We show that the asymptotic solution is an infinite sum of saddle point contributions that depend on the value of $x$. In region $\mathcal{D}_1$, only the upper saddle points $z_s^+$ contribute. In region $\mathcal{D}_2$, only the lower saddle points $z_s^-$ contribute. In region $\mathcal{D}_3$, both sets of saddle points, $z_s^+$ and $z_s^-$, contribute.

Figure \ref{fig:dAistokstruc} shows the regions $\mathcal{D}_1$, $\mathcal{D}_2$, and $\mathcal{D}_3$, separated by active Stokes curves. The curves $\mathcal{S}_{s,s}^{+,-}$ originate at $x = 0$, while $\mathcal{S}_{s,s+1}^{-,+}$ originate at $x = -4$. These curves intersect at two Stokes crossing points, located at $x \approx -2 \pm 3.018\mathrm{i}$. The Stokes curves $\mathcal{S}_{s,s}^{+,-}$ and $\mathcal{S}_{s,s+1}^{-,+}$ become inactive at the Stokes crossing points. Additional Stokes curves $\mathcal{S}_{s,s+1}^{-,-}$ and $\mathcal{S}_{s,s+1}^{+,+}$ extend vertically from the crossing points but also become inactive there.

Figure \ref{fig:sdreg} shows typical steepest descent schematics in regions $\mathcal{D}_1$, $\mathcal{D}_2$, and $\mathcal{D}_3$. This analysis reveals that the solution can be written as
\begin{equation}
\label{eqn:dtransser}
    y \sim  \sum_{s=-\infty}^{\infty} \left( c_{s}^-y_s^- +  c_{s}^+y_s^+ \right) \quad \text{as} \quad \epsilon \to 0 \, ,
\end{equation}
where $c_s^+ = 1$ and $c_s^- = 0$ if $x \in \mathcal{D}_1$, $c_s^+ = 1$ and $c_s^- = 1$ if $x \in \mathcal{D}_2$, and $c_s^+ = 0$ and $c_s^- = 1$ if $x \in \mathcal{D}_3$. These coefficients change across Stokes curves, which separate $\mathcal{D}_1$, $\mathcal{D}_2$ and $\mathcal{D}_3$. 

In Section \ref{s:SDcurves}, we will demonstrate how the steepest descent contour varies between these regions, to explain the switching behaviour observed in Figures \ref{fig:dAistokstruc} and \ref{fig:sdreg}. 

\subsubsection{Steepest descent curves}\label{s:SDcurves}

\begin{figure}[tb!]
\centering
   \subfloat[Stokes structure]{\includegraphics[height=0.35\textwidth]{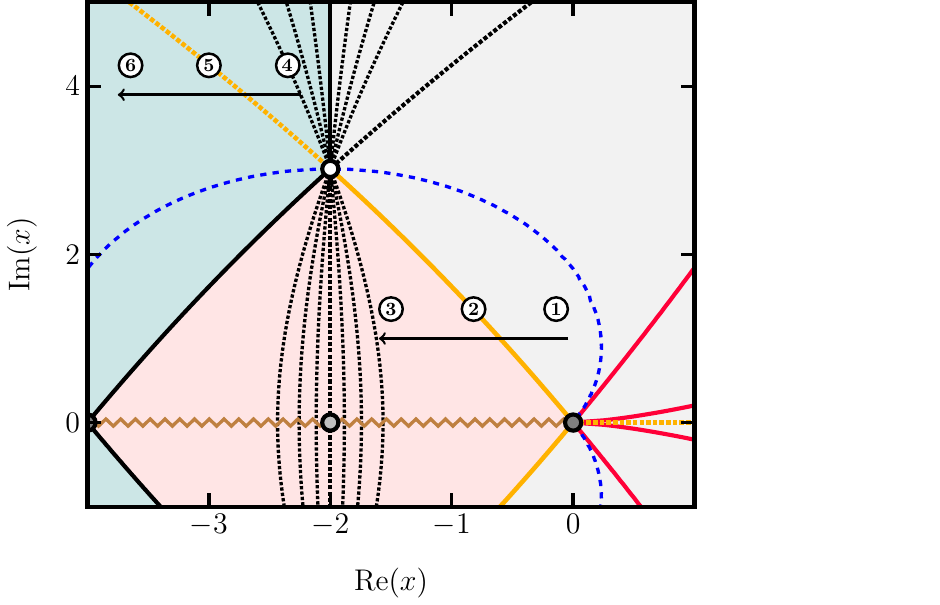}} 
   
   \subfloat[Steepest descent contour \ding{172}]{\includegraphics[width=0.31\textwidth]{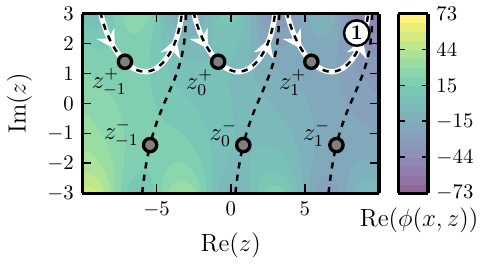}} 
   \subfloat[Steepest descent contour \ding{173}]{\includegraphics[width=0.31\textwidth]{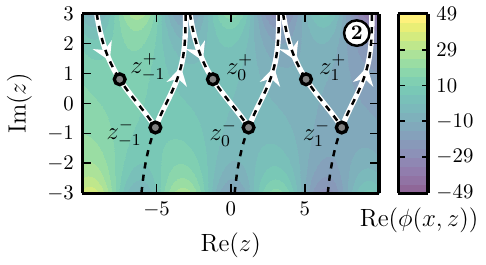}} 
   \subfloat[Steepest descent contour \ding{174}]{\includegraphics[width=0.31\textwidth]{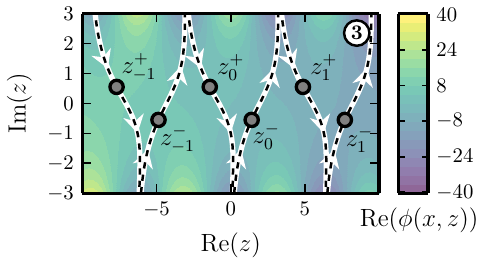}} 

   \subfloat[Steepest descent contour \ding{175}]{\includegraphics[width=0.31\textwidth]{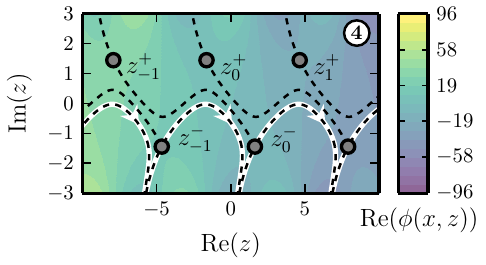}} 
   \subfloat[Steepest descent contour \ding{176}]{\includegraphics[width=0.31\textwidth]{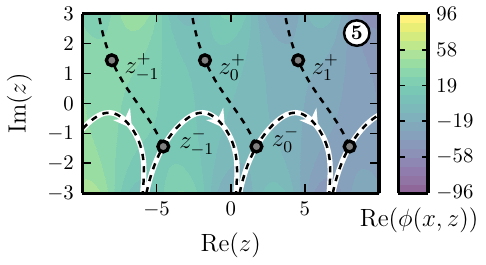}} 
   \subfloat[Steepest descent contour \ding{177}]{\includegraphics[width=0.31\textwidth]{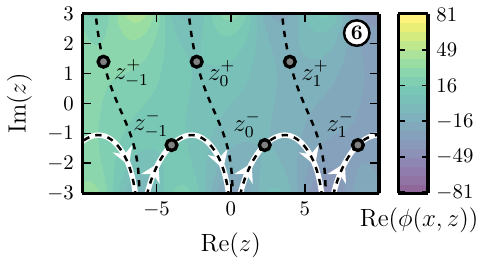}} 
    \caption{Schematics of Stokes switching for the Stokes curves $\mathcal{S}_{s,s}^{-,+}$ with $\sigma = 1$. (a) Stokes structure schematic. Showing Stokes curves $\mathcal{S}_{s,s}^{-,+}$ (yellow), each emanate from the turning point $x = 0$. The legend otherwise matches that of Figure \ref{fig:dAistokstruc} with additional inactive Stokes curves (dotted yellow and black). Points \ding{172}--\ding{177} correspond to the steepest descent schematics shown in (b)--(g). (b)--(g) Steepest descent schematics, showing the integration path (white) and saddle points $z_s^\pm$ (circles).}
    \label{fig:s1sd}
\end{figure}

\paragraph{Stokes Switching Across $\mathcal{S}_{s,s}^{+,-}$:}

In Figure \ref{fig:s1sd}, we show how the steepest descent contour changes as the Stokes curves $\mathcal{S}_{s,s}^{+,-}$ are crossed. Following the labelled path from \ding{172} to \ding{173} to \ding{174}, we observe that as the Stokes curves are crossed, each saddle $z_s^+$ in the upper row switches on a saddle $z_s^-$ in the lower row. Hence, additional 
contributions from the $z_s^-$ saddles emerge in the asymptotic solution as $x$ moves from $\mathcal{D}_1$ to $\mathcal{D}_2$.

Following the path from \ding{175} to \ding{176} to \ding{177}, we see that the Stokes curves are inactive here. Although $z_s^+$ and $z_{s}^-$ have equal phase at \ding{176}, the steepest descent contour does not change, so no Stokes switching occurs. 

\paragraph{Stokes Switching Across $\mathcal{S}_{s,s+1}^{-,+}$:}

\begin{figure}[tb!]
    \centering
    \subfloat[Stokes structure]{\includegraphics[height=0.35\textwidth]{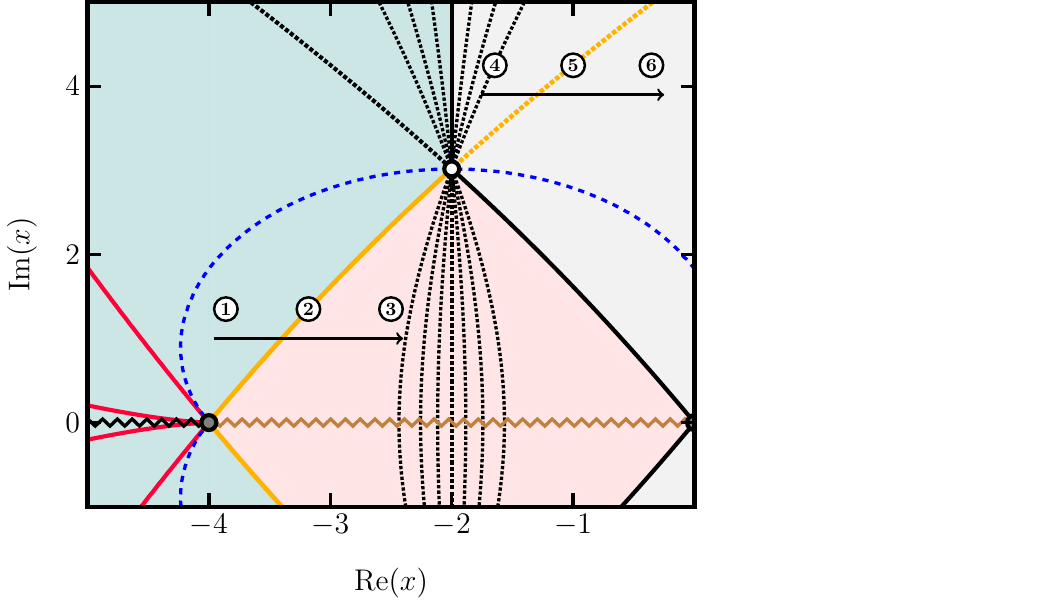}} 
    
   \subfloat[Steepest descent contour \ding{172}]{\includegraphics[width=0.31\textwidth]{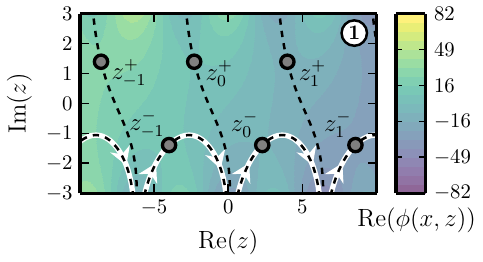}} 
   \subfloat[Steepest descent contour \ding{173}]{\includegraphics[width=0.31\textwidth]{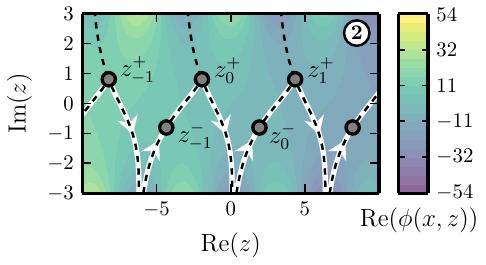}} 
   \subfloat[Steepest descent contour \ding{174}]{\includegraphics[width=0.31\textwidth]{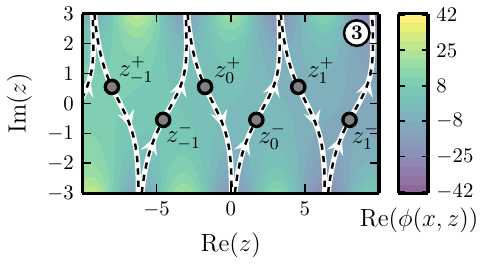}} 
   
   \subfloat[Steepest descent contour \ding{175}]{\includegraphics[width=0.31\textwidth]{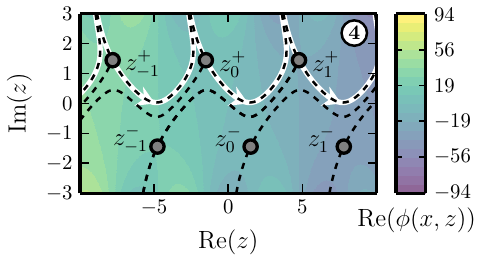}} 
   \subfloat[Steepest descent contour \ding{176}]{\includegraphics[width=0.31\textwidth]{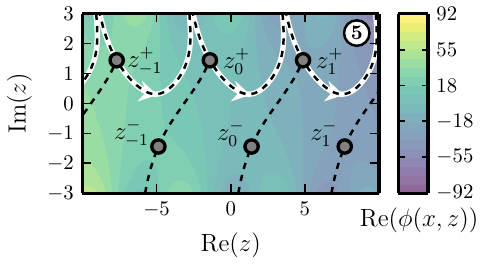}} 
   \subfloat[Steepest descent contour \ding{177}]{\includegraphics[width=0.31\textwidth]{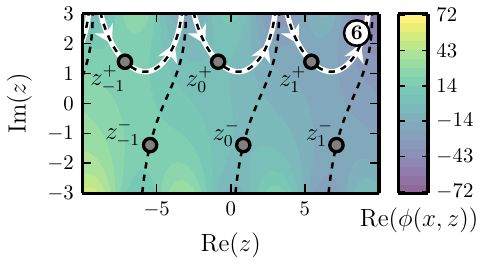}}
    \caption{
    Schematics of Stokes switching for the Stokes curves $\mathcal{S}_{s,s+1}^{-,+}$ with $\sigma = 1$. (a) Stokes structure schematic. Showing Stokes curves $\mathcal{S}_{s,s+1}^{-,+}$ (yellow), each emanate from the turning point $x = -4$. The legend otherwise matches that of Figure \ref{fig:dAistokstruc} with additional inactive Stokes curves (dotted yellow and black). Points \ding{172}--\ding{177} correspond to the steepest descent schematics shown in (b)--(g). (b)--(g) Steepest descent schematics, showing the integration path (white) and saddle points $z_s^\pm$ (circles).}
    \label{fig:s2sd}
\end{figure}

In Figure \ref{fig:s2sd}, we show how the steepest descent contour changes as the Stokes curves $\mathcal{S}_{s,s+1}^{-,+}$ are crossed. Following the labelled path from \ding{172} to \ding{173} to \ding{174}, we observe that as the Stokes curves are crossed, each saddle $z_s^-$ in the lower row switches on a saddle $z_{s+1}^+$ in the upper row. Hence, additional 
contributions from the $z_s^+$ saddles emerge in the asymptotic solution as $x$ moves from $\mathcal{D}_2$ to $\mathcal{D}_3$.

Following the path from \ding{175} to \ding{176} to \ding{177}, we see that the Stokes curves are inactive here. Although $z_s^-$ and $z_{s+1}^+$ have equal phase at \ding{176}, the steepest descent contour does not change, so no Stokes switching occurs. 

\paragraph{Stokes Switching Across $\mathcal{S}_{s,s+1}^{+,+}$ and $\mathcal{S}_{s,s+1}^{-,-}$:}

\begin{figure}[tb!]
    \centering
    \subfloat[Stokes structure]{\includegraphics[height=0.35\textwidth]{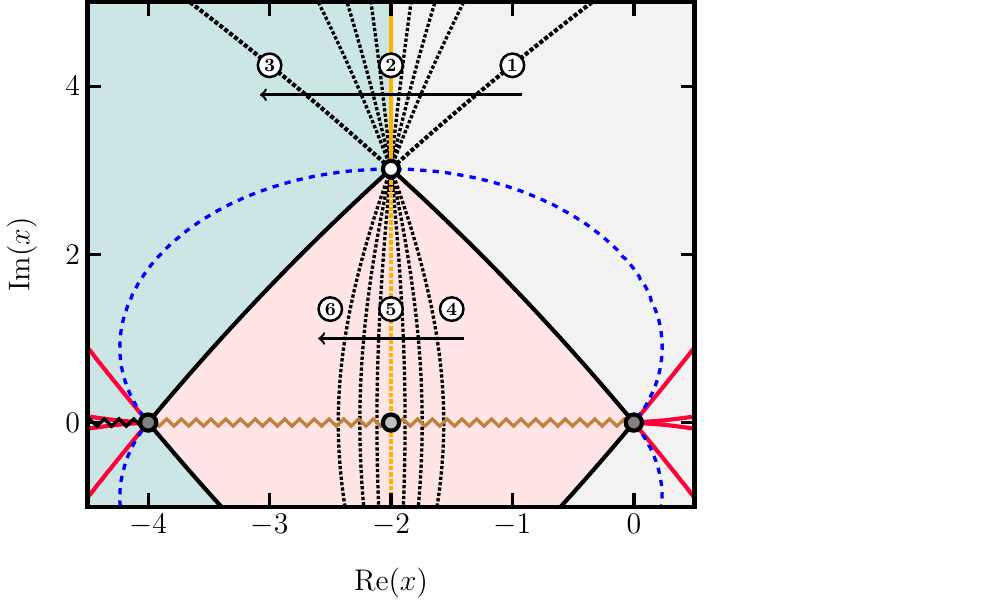}} 
    
   \subfloat[Steepest descent contour \ding{172}]{\includegraphics[width=0.31\textwidth]{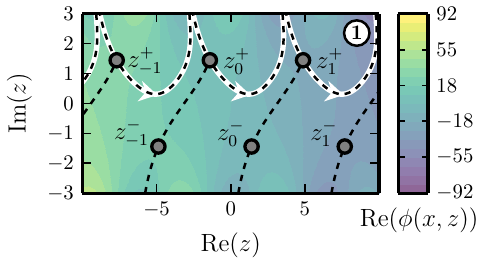}} 
   \subfloat[Steepest descent contour \ding{173}]{\includegraphics[width=0.31\textwidth]{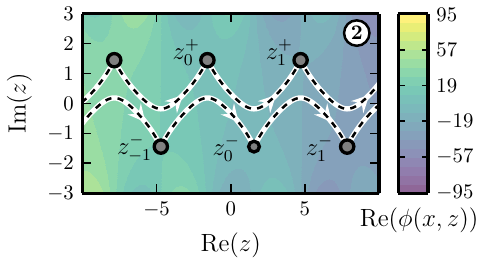}}
   \subfloat[Steepest descent contour \ding{174}]{\includegraphics[width=0.31\textwidth]{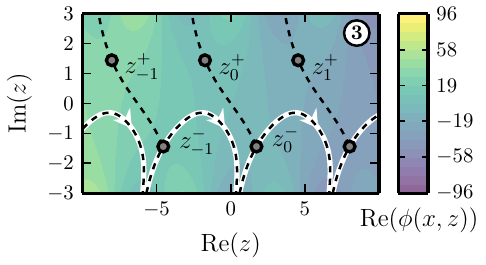}} 
   
   \subfloat[Steepest descent contour \ding{175}]{\includegraphics[width=0.31\textwidth]{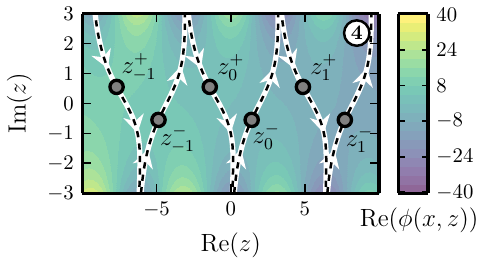}} 
   \subfloat[Steepest descent contour \ding{176}]{\includegraphics[width=0.31\textwidth]{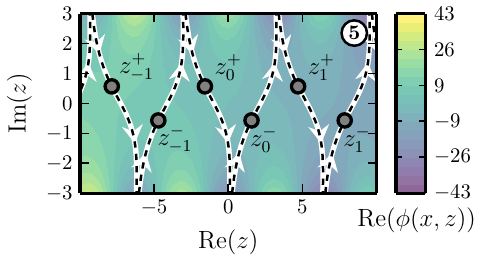}}
    \subfloat[Steepest descent contour \ding{177}]{\includegraphics[width=0.31\textwidth]{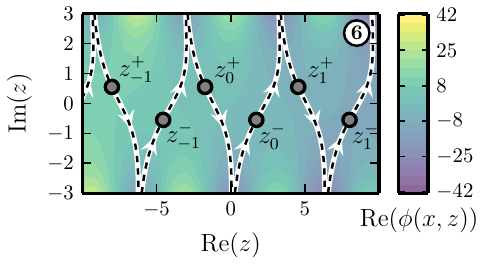}} 
    \caption{Schematics of Stokes switching for the Stokes curves $\mathcal{S}_{s,s+1}^{+,+}$ and $\mathcal{S}_{s,s+1}^{-,-}$ with $\sigma = 1$. (a) Stokes structure schematic. Showing Stokes curves $\mathcal{S}_{s,s+1}^{+,+}$ and $\mathcal{S}_{s,s+1}^{-,-}$ (yellow), each emanate from the virtual turning point $x = -2$. The legend otherwise matches that of Figure \ref{fig:dAistokstruc} with additional inactive Stokes curves (dotted yellow and black). Points \ding{172}--\ding{177} correspond to the steepest descent schematics shown in (b)--(g). (b)--(g) Steepest descent schematics, showing the integration path (white) and saddle points $z_s^\pm$ (circles).}
    \label{fig:s3sd}
\end{figure}

In Figure \ref{fig:s3sd}, we show how the steepest descent contour changes as the Stokes curves $\mathcal{S}_{s,s+1}^{-,+}$ are crossed. Following the labelled path from \ding{172} to \ding{173} to \ding{174}, we observe that as the Stokes curves are crossed, each saddle $z_s^+$ in the upper row switches off and each saddle $z_{s}^-$ in the lower row switches on simultaneously. Hence, contributions from the $z_s^-$ saddles appear and contributions from the $z_s^+$ saddles disappear in the asymptotic solution as $x$ moves from $\mathcal{D}_1$ to $\mathcal{D}_2$. This unusual switching behaviour, in which all saddle contributions switch on or off simultaneously, is caused by the integral endpoint at $z = -\infty$.

Following the path from \ding{175} to \ding{176} to \ding{177}, the Stokes curves are inactive here. Although pairs $z_s^-$ and $z_{s+1}^-$, and $z_{s}^+$ and $z_{s+1}^+$ have equal phase at \ding{176}, the steepest descent contour does not change, so no Stokes switching occurs. 

\paragraph{Higher-Order Stokes Switching Across $\mathcal{H}_{s,s+1,s+1}^{-,+,-}$ and $\mathcal{H}_{s,s,s+1}^{+,-,+}$:}

Figures \ref{fig:s1sd}, \ref{fig:s2sd}, and \ref{fig:s3sd} illustrate how saddle point adjacency changes upon crossing the higher-order Stokes curves $\mathcal{H}_{s,s+1,s+1}^{-,+,-}$ and $\mathcal{H}_{s,s,s+1}^{+,-,+}$. Stokes switching only occurs between adjacent saddles. Inside the region bounded by these curves, $z_s^+$ is adjacent to $z_{s}^-$ and $z_{s-1}^-$, while $z_s^-$ is adjacent to $z_{s}^+$ and $z_{s+1}^+$, allowing switching across $\mathcal{S}_{s,s}^{+,-}$ and $\mathcal{S}_{s,s+1}^{-,+}$. Outside this region, $z_s^+$ is adjacent to $z_{s+1}^+$ and $z_s^-$ is adjacent to $z_{s+1}^-$, enabling switching across $\mathcal{S}_{s,s+1}^{+,+}$ and $\mathcal{S}_{s,s+1}^{-,-}$.

\paragraph{Branch cuts:}

The branch cuts do not affect which saddles contribute to the asymptotic solution. The logarithmic branch cut along real $x < -4$, maps $z_s^- \to z_{s+1}^-$; since it is within $\mathcal{D}_3$, all saddles in the lower row contribute. The branch cut along real $-4 < x < 0$, maps $z_s^- \leftrightarrow z_s^+$; since it is within $\mathcal{D}_2$, all saddles contribute. The branch cuts therefore relabel contributions but do not alter the asymptotic solution.

\subsection{Accumulation of Stokes and anti-Stokes curves}

\begin{figure}[tb!]
        \centering
    \includegraphics[width=0.9\textwidth]{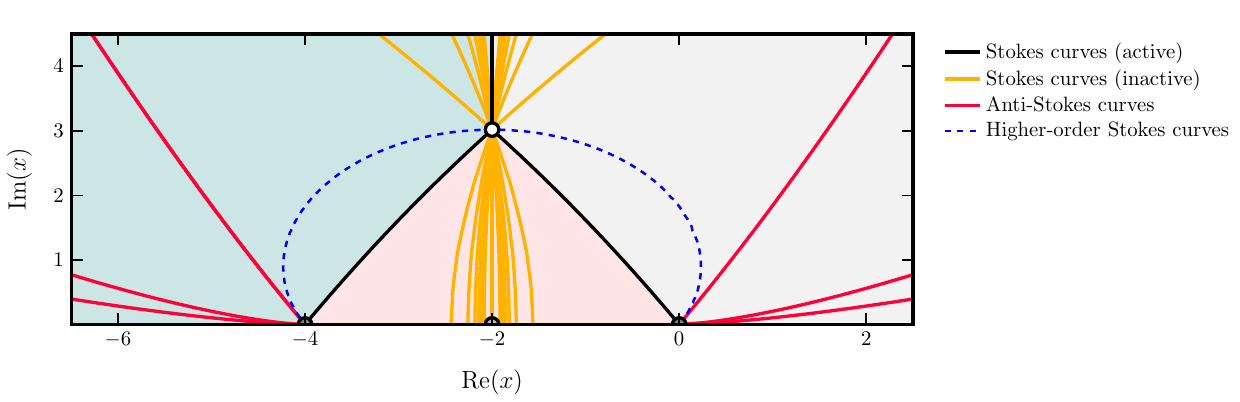}
    \caption{Schematic of Stokes and anti-Stokes curve accumulations for the advance-delay Airy equation \eqref{eqn:soadvAi} for $\sigma = 1$. Anti-Stokes curves accumulate on the real axis for $x > 0$ and $x < -4$, while inactive Stokes curves accumulate toward $\mathrm{Re}(x) = -2$.}
    \label{fig:accum}
\end{figure}

The steepest descent analysis fully describes the asymptotic solutions to the advance-delay Airy equation \eqref{eqn:soadvAi}. In this section, we comment on an unusual feature, not typically seen in solutions to linear differential equations. The solution contains an infinite number of anti-Stokes curves accumulating towards the real axis and (inactive) Stokes curves accumulating towards $\mathrm{Re}(x) = -2$. These accumulations are illustrated in Figure \ref{fig:accum}.

For $\mathrm{Re}(x) \leq 4$, the anti-Stokes curves $\mathcal{A}_{s,s+j}^{+,-}$ with $j \geq 1$ accumulate onto the real axis, and for $\mathrm{Re}(x) \geq 0$, the curves $\mathcal{A}_{s,s+j}^{-,+}$ with $j \geq 1$ accumulate similarly. As the real axis is approached, infinitely many anti-Stokes curves cause saddle contributions to switch from exponentially large to small. Although these curves do not change the form of the solution \eqref{eqn:dtransser}, they induce a rapid shift in the dominant balance of contributions near the real axis.

The inactive Stokes curves $\mathcal{S}_{s,s+j}^{-,+}$ and $\mathcal{S}_{s,s+j}^{+,-}$ with $j \geq 1$ accumulate toward the line $\mathrm{Re}(x) = -2$. Between the Stokes crossing points, the Stokes curves are $\mathcal{S}_{s,s+j}^{-,+}$ with $j \geq 1$ for $\mathrm{Re}(x) < -2$, while the curves are $\mathcal{S}_{s,s+j}^{+,-}$ with $j \geq 1$ for $\mathrm{Re}(x) > -2$; the reverse is true outside of the Stokes crossing points. 

Each inactive Stokes curve corresponds to a pair of saddles with equal phase, but no Stokes switching occurs. As shown in Figure \ref{fig:accum}, approaching the line $\mathrm{Re}(x) = -2$ reveals a sequence of inactive Stokes curves connecting $z_s^-$ to $z_{s+2}^+$, $z_{s+3}^+$, and so on. Although these curves do not influence the present solution of the discrete Airy equation, the accumulation of infinitely many (even inactive) Stokes curves is an unusual feature in linear problems.

These unusual accumulations occur because the asymptotic solution \eqref{eqn:dtransser} contains infinitely many saddle contributions, leading to infinitely many Stokes switching interactions and the resulting curve accumulations. In continuous differential equations, such behaviour is typically linked to nonlinearity, where the transseries has infinitely many terms \cite{chapman2007}. Here, however, it arises in a linear discrete system from the 
$2\pi$-periodicity of the saddle locations, a generic feature of discrete problems (see, eg. \cite{joshi2015,king2001, lustri2020,moston2022}). In these previous studies, only the Stokes curves associated with the dominant contributions are considered and therefore the accumulations are not identified. Thus, unlike continuous systems, the observed accumulation of Stokes and anti-Stokes curves is not due to nonlinearity, but is instead due to the discretization.

\subsection{Varying \texorpdfstring{$\sigma$}{sigma}}
\label{sect:difsig}

\begin{figure}[tb!]
\subfloat[$\sigma = 0.75$]{
\centering
\includegraphics[width=0.3\textwidth]{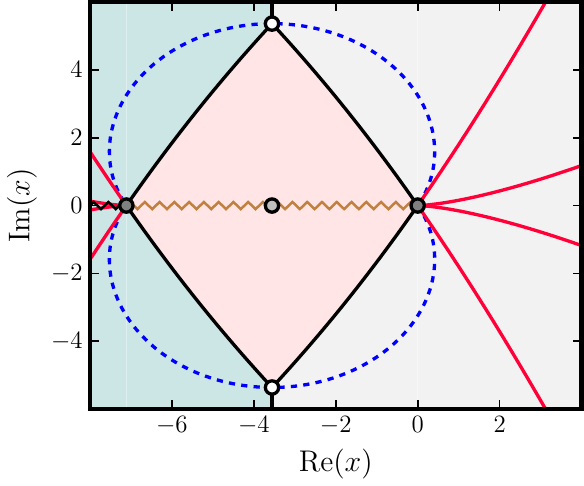}}~
\subfloat[$\sigma = 1$]{
\centering
\includegraphics[width=0.3\textwidth]{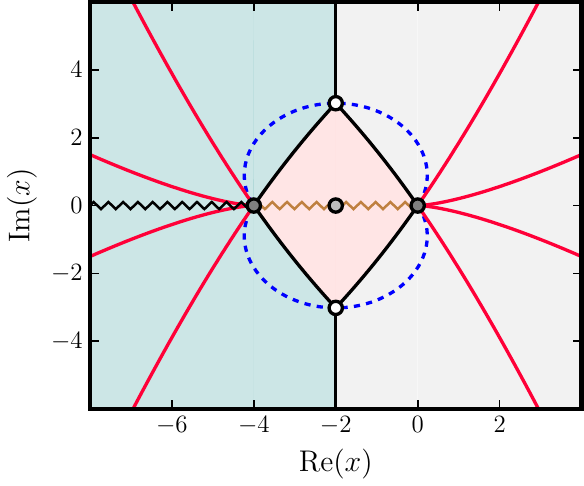}}~
\subfloat[$\sigma = 1.25$]{
\centering
\includegraphics[width=0.3\textwidth]{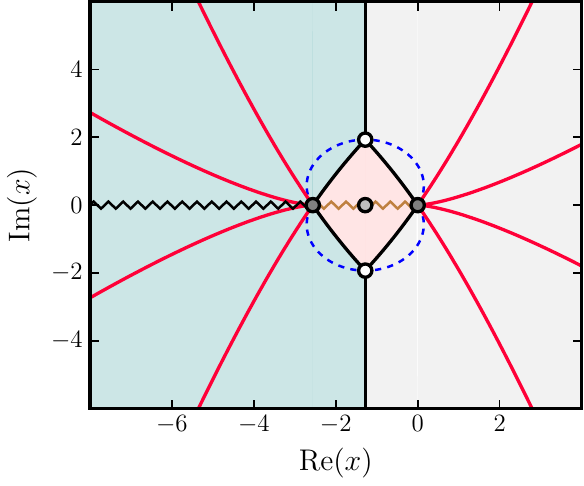}}
\caption{Schematic showing how $|\sigma|$ affects the Stokes structure. The legend is identical to Figure \ref{fig:dAistokstruc}. Panels (a)–(c) show the Stokes structure for varying $|\sigma|$ with $\Arg(\sigma)=0$. The turning point $x=0$ remains fixed, while the structure scales with $1/|\sigma|^2$.}
    \label{fig:radstok}
\end{figure}

\begin{figure}[tb!]
\subfloat[$\Arg\{\sigma\} = -\pi/12$]{
\centering
\includegraphics[width=0.23\textwidth]{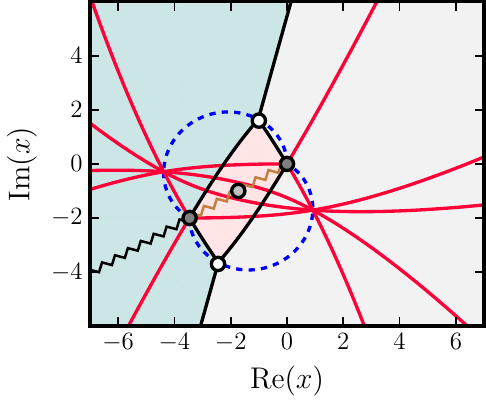}} 
\subfloat[$\Arg\{\sigma\} = 0$]{
\centering
\includegraphics[width=0.23\textwidth]{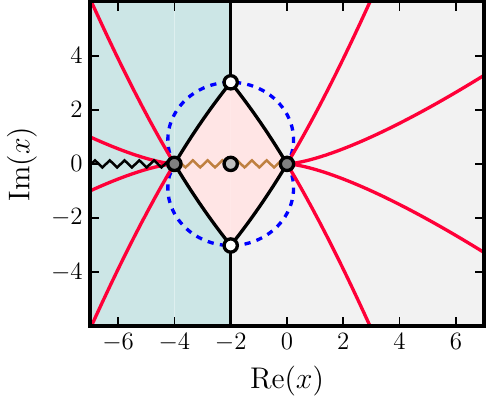}} 
\subfloat[$\Arg\{\sigma\} = \pi/12$]{
\centering
\includegraphics[width=0.23\textwidth]{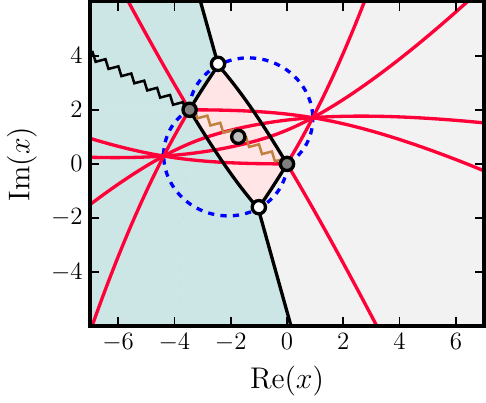}} ~
\subfloat[$\Arg\{\sigma\} = \pi/6$]{
\centering
\includegraphics[width=0.23\textwidth]{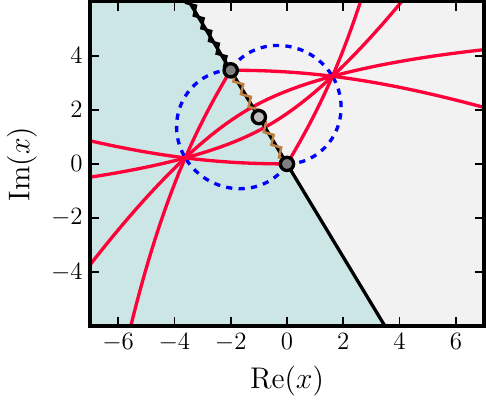}}

\subfloat[$\Arg\{\sigma\} = \pi/4$]{
\centering
\includegraphics[width=0.23\textwidth]{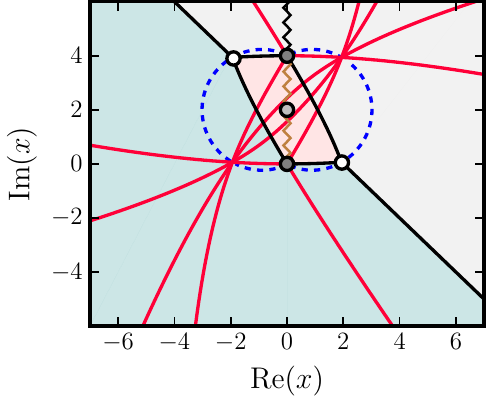}} 
\subfloat[$\Arg\{\sigma\} = \pi/3$]{
\centering
\includegraphics[width=0.23\textwidth]{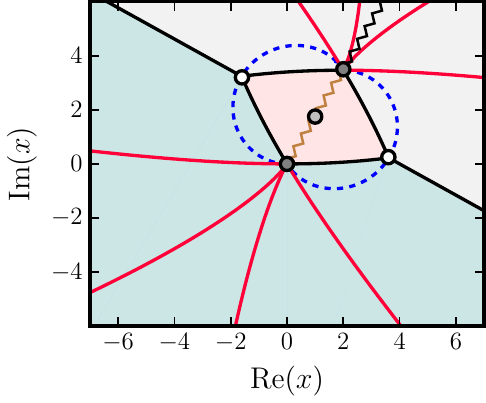}} 
\subfloat[$\Arg\{\sigma\} = 5\pi/12$]{
\centering
\includegraphics[width=0.23\textwidth]{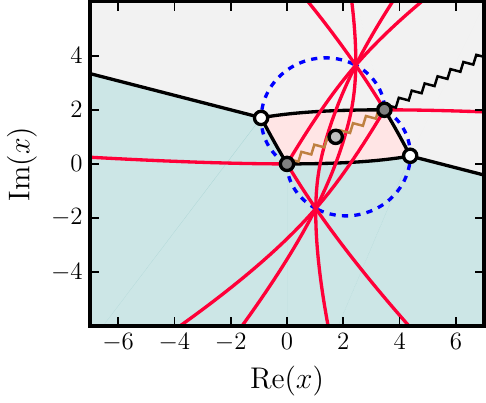}} 
\subfloat[$\Arg\{\sigma\} = \pi/2$]{
\centering
\includegraphics[width=0.23\textwidth]{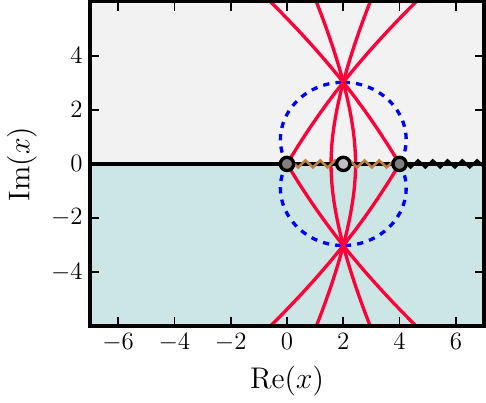}}
\caption{Schematic showing how $\Arg(\sigma)$ affects the Stokes structure. The legend is identical to Figure \ref{fig:dAistokstruc}. Panels (a)–(h) show the Stokes structure for varying $\Arg(\sigma)$ with $|\sigma|=1$. The turning point $x=0$ remains fixed. As $\Arg(\sigma)$ varies, key features of the Stokes structure deform and rotate. The regions $\mathcal{D}_1$, $\mathcal{D}_2$ and $\mathcal{D}_3$ change accordingly.}
\label{fig:sigstok}
\end{figure}

In Section \ref{stepdescsigone}, we present the Stokes structure and asymptotic solutions of the advance-delay Airy equation \eqref{eqn:soadvAi} for $\sigma = 1$. Allowing $\sigma$ to vary allows the spatial step $h$ to change independently of the small parameter $\epsilon$, including the use of complex spatial steps when $\sigma$ is complex. This is a useful generalisation as many special functions satisfy difference equations from which their asymptotic properties in complex directions can be determined \cite{huang2019, huang2020,joshi2015, li2020, wang2012}. This generalisation significantly alters the Stokes structure. In this section, we examine how the Stokes structure changes as $|\sigma|$ and $\Arg(\sigma)$ vary.

Figure \ref{fig:radstok} illustrates how $|\sigma|$ affects the Stokes structure. The turning point at $x = 0$ remains fixed, while the other turning and virtual turning points lie at $x = -4/\sigma^2$ and $x = -2/\sigma^2$, respectively. Since active Stokes curves emerge from these points, the Stokes structure scales with $1/|\sigma|^2$, and the regions $\mathcal{D}_1$, $\mathcal{D}_2$, and $\mathcal{D}_3$ scale accordingly. As $|\sigma|\to 0$ the Stokes structure near $x = 0$ approaches that of the continuous Airy equation \eqref{eqn:spAi}.

While changing $|\sigma|$ results in a straightforward scaling of the Stokes structure, altering $\Arg(\sigma)$ has a more significant impact. Figure \ref{fig:sigstok} illustrates how $\Arg(\sigma)$ affects the Stokes structure. The turning points lie along the ray $\Arg(x) = \pi - 2\Arg(\sigma)$, while the Stokes crossing points lie along the rays $\Arg(x + 2/\sigma^2) = \pm (\pi/2 + \Arg(\sigma))$. As $\Arg(\sigma)$ varies, key features of the Stokes structure deform and rotate, and the geometry of the regions $\mathcal{D}_1$, $\mathcal{D}_2$, and $\mathcal{D}_3$ changes accordingly. For $\Arg(\sigma) = (2 + 4n)\pi/2$, with $n \in \{0, 1, 2, 3, 4, 5\}$, each Stokes crossing point coalesces with a turning point. The active Stokes curves $\mathcal{S}_{s,s}^{+,-}$ and $\mathcal{S}_{s,s+1}^{-,+}$ then coincide, eliminating region $\mathcal{D}_3$. In these cases, only $\mathcal{D}_1$ and $\mathcal{D}_2$ remain, and there is no region in which all contributions are simultaneously present.

\section{Comparison Of Results}
    \label{sect:NUM}

\subsection{Numerical comparison}

To validate our asymptotic solutions \eqref{eqn:dtransser} of the advance-delay Airy equation \eqref{eqn:soadvAi}, we compare them with numerical solutions of the discrete Airy equation \eqref{eqn:latAi}. Solutions of \eqref{eqn:soadvAi} can be written as the matrix system $M \mathbf{y} = \mathbf{0}$, where the only nonzero entries are given by $M_{m-1,m} = 1/\sigma^2$, $M_{m,m} = -2/\sigma^2 - x_{m}$, $M_{m+1,m} = 1/\sigma^2$, and $\mathbf{y}_m = y_m$.

We seek decaying solutions such that $y_m \to 0$ as $|x_m| \to \infty$. To approximate the infinite-dimensional system, we truncate the domain to a finite interval $x_m \in [x_{M_1}, x_{M_2}]$, with $x_{M_1}$ and $x_{M_2}$ chosen so that the boundary values are negligibly small. We therefore impose $y_{M_1} = y_{M_2} = 0$, setting a value for $y_0$, gives a solvable system. To ensure accuracy, we solve the system on progressively larger domains until the solution converges.

Figure \ref{fig:dAiNSol} shows numerical solutions to the discrete Airy equation \eqref{eqn:latAi} for real $x_m \in \mathbb{R}$ with $\sigma = 1$ and various values of $\epsilon$. The scheme is implemented with $y_0 = 1$ at $x_0 = -2$, and solutions are normalised to have a maximum value of 1 for visual clarity. The solution is oscillatory between $x \approx -4$ and $x \approx 0$, and decays exponentially outside this region. Near $x = -4$ and $x = 0$, the envelope follows an Airy profile. These features are consistent with the asymptotic prediction that the solution locally resembles Airy function behavior near each turning point.

For certain values of $\sigma$, $\epsilon$, $h$, and $x_m$, the asymptotic solution simplifies, enabling direct comparison with numerical results. If $(x_m+2/\sigma^2)/(\sigma\epsilon)\in\mathbb{Z}$ for all $m$, then the terms $y_s^-$ and $y_p^-$ are identical up to a scalar multiple for all $s$ and $p$, and similarly for $y_s^+$ and $y_p^+$. In this case, the asymptotic solution \eqref{eqn:dtransser} reduces to
\begin{equation}
\label{transnsns2}
y \sim \begin{cases} cy_{0}^{+} & \text{for $x\in\mathcal{D}_{1}$}  \\ cy_{0}^{-} & \text{for $x\in\mathcal{D}_2$} \\ cy_0^-+cy_0^+ & \text{for $x\in\mathcal{D}_3$ }\end{cases} \, .
\end{equation}

\begin{figure}[tb!]
    \centering
    \subfloat[$\sigma=1$ and $\epsilon=0.05$]{\includegraphics[width=0.49\textwidth]{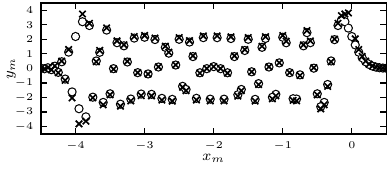}} 
    \subfloat[$\sigma=1$ and $\epsilon=0.025$]{\includegraphics[width=0.49\textwidth]{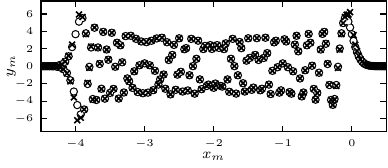}} 

    \subfloat[$\sigma=1$ and $\epsilon=0.0125$]{\includegraphics[width=0.49\textwidth]{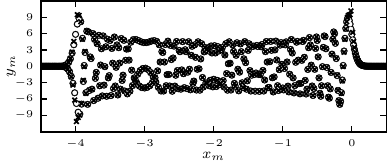}} 
    \subfloat[$\sigma=1$ and $\epsilon=0.005$]{\includegraphics[width=0.49\textwidth]{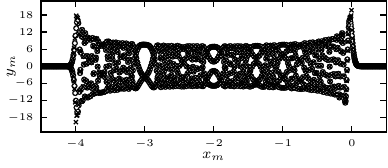}} 
    \caption{Asymptotic solutions \eqref{eqn:dtransser} of the advance-delay Airy equation \eqref{eqn:soadvAi} and numerical solutions of the discrete Airy equation \eqref{eqn:spAi}, for several values of $(x_m+2/\sigma^2)/(\sigma\epsilon)\in\mathbb{Z}$. The continuous asymptotic solution is sampled at discrete grid points $x_m$. Numerical results (circles) and asymptotic results (crosses) are in agreement, with the accuracy improving as $\epsilon \to 0$. In Figures (b), (c), and (d), the solutions are visually indistinguishable except near the turning points.}
    \label{fig:numasy}
\end{figure}

Figure \ref{fig:numasy} shows numerical solutions of the discrete Airy equation \eqref{eqn:latAi} compared to the asymptotic solution \eqref{transnsns2}. The numerical scheme uses the initial condition $y_0 = y(x_0)$ at $x_0 = -2$, where $y(x_0)$ is computed from \eqref{transnsns2} with $c = 1$. As $\epsilon \to 0$, the agreement between numerical and asymptotic solutions improves.

\begin{figure}[tb!]
    \centering
    \subfloat[$\Arg(\sigma)=0$]{\includegraphics[width=0.49\textwidth]{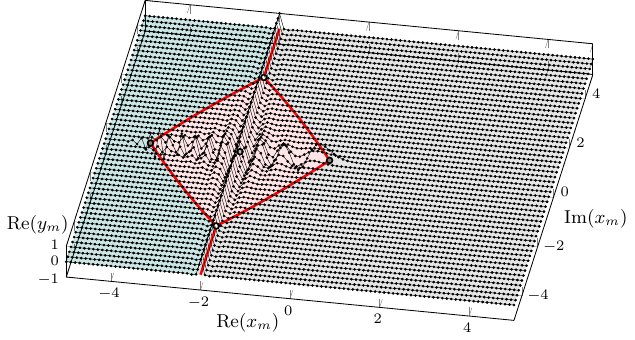}} 
    \subfloat[$\Arg(\sigma)=\pi/12$]{\includegraphics[width=0.49\textwidth]{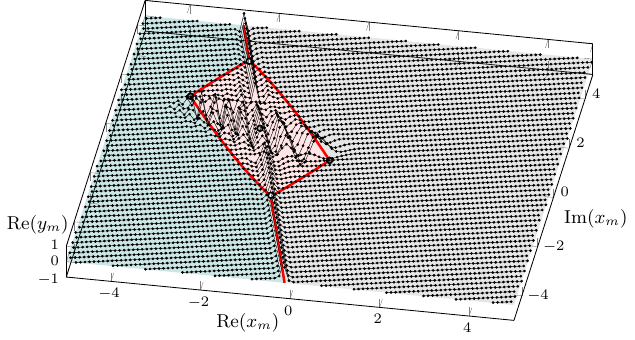}} 

    \subfloat[$\Arg(\sigma)=2\pi/12$]{\includegraphics[width=0.49\textwidth]{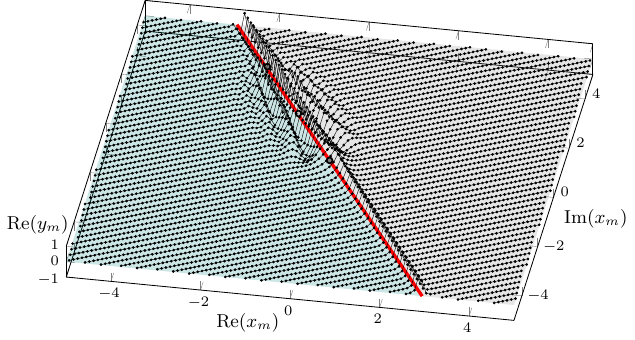}} 
    \subfloat[$\Arg(\sigma)=3\pi/12$]{\includegraphics[width=0.49\textwidth]{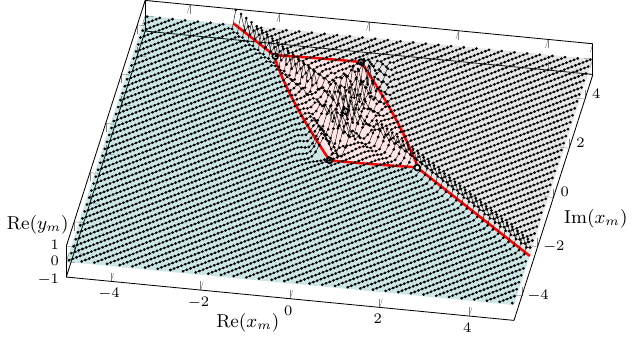}} 
    \caption{Comparison of the Stokes structure of our asymptotic solutions \eqref{eqn:dtransser} to the advance-delay Airy equation \eqref{eqn:soadvAi} with numerical solutions of the discrete Airy equation \eqref{eqn:spAi}, for several values of $\Arg(\sigma)$ with $\epsilon = 0.125$ and $|\sigma| = 1$. Numerical solutions (black), are normalized along each line to have a maximum value of 1. The asymptotic and numerical results are in close agreement, showing oscillations with a slowly varying envelope in $\mathcal{D}_3$ and exponential decay in $\mathcal{D}_1$ and $\mathcal{D}_2$.}
    \label{fig:numasy2}
\end{figure}

Figure \ref{fig:numasy2} shows numerical solutions of the discrete Airy equation \eqref{eqn:latAi} compared to the Stokes structure of our asymptotic solutions \eqref{eqn:dtransser} to the advance-delay Airy equation \eqref{eqn:soadvAi}. Comparisons are made along several lines $x_m$ in the complex plane for several values of $\Arg(\sigma)$ with $\epsilon = 0.125$ and $|\sigma| = 1$. The numerical scheme is implemented with the initial condition $y_0 = 1$ at the corresponding location $x_0 = -2/\sigma^2$, and $y_m = 0$ at the end-points of the region. This does not correspond to the specific solution from section \ref{stepdescsigone}, which has a different boundary condition as $|\mathrm{Re}(x_m)| \to \infty$ in the upper half-complex plane; however, the two solutions must exhibit the same Stokes structure. These simulations therefore do provide a numerical validation of the predicted Stokes structure from the steepest descent analysis.

The Stokes structure predicted through asymptotic analysis matches the numerical results, showing oscillatory behavior with a slowly varying envelope between the turning points in the enclosed region $\mathcal{D}_3$, and exponential decay in the outer regions $\mathcal{D}_1$ and $\mathcal{D}_2$.

\subsection{Comparison with existing literature}
\label{sect:CWL}

In our study of the discrete Airy equation \eqref{eqn:latAi}, we determined the transseries solution \eqref{eqn:dtransser}, whose coefficients vary across the regions $\mathcal{D}_1$, $\mathcal{D}_2$, and $\mathcal{D}_3$. These regions are bounded by active Stokes curves. The coefficient changes arise from the Stokes phenomenon as these curves are crossed. The Stokes curves originate at two turning points and a virtual turning point, and truncate at two Stokes crossing points. At the Stokes crossing points, higher-order Stokes curves emerge, across which higher-order Stokes switching occurs, truncating the active Stokes curves.

A key novelty of our work is the identification of a virtual turning point at $x=-2/\sigma^2$ and the associated higher-order Stokes phenomenon. These features were not observed in previous studies \cite{burr2000,cao2014,fedotov2019,geronimo2009,kashani2016,li2020,wang2003,wang2005,wong1992b}, which focus on solutions along lines through the origin. Detecting the virtual turning point requires analytic continuation away from such lines, and hence the Stokes structure in the complex plane was not considered. Although \cite{kashani2016} presents a related Stokes structure for a different discrete equation, the Stokes curves in that case do not intersect, and thus no virtual turning point or higher-order Stokes switching is observed.

The turning points at $x=-4/\sigma^2$ and $x=0$, agree with those identified using direct series methods \cite{cao2014,li2020,wang2003,wang2005,wong1992b} and WKB methods \cite{burr2000, fedotov2019, geronimo2009, kashani2016}. Near the turning points, the solution \eqref{eqn:dtransser} follows an Airy function envelope, away from these points, the solution deviates from Airy function behavior, in agreement with these prior findings.

Near the turning points, the geometry of the active Stokes curves resembles that of the Airy function, consistent with studies \cite{burr2000,fedotov2022, fedotov2019,fedotov2021,kashani2016}. Away from these points, the Stokes structure deviates due to differences in the singulants of the advance-delay Airy equation \eqref{eqn:soadvAi} and the Airy equation \eqref{eqn:spAi}. As previous studies consider different discrete equations, a direct comparison of the Stokes structure is not possible, however, deviation from Airy function behavior is consistently observed \cite{fedotov2022,fedotov2019,fedotov2021,kashani2016}.

In Section \ref{sect:difsig}, we examine how the Stokes structure depends on both $\Arg{\sigma}$ and $|\sigma|$ for complex step sizes $h$. Previous studies \cite{cao2014,li2020,wang2003,wang2005,wong1992b} developed different asymptotic constructions based on relative positioning of the turning points, consistent with our observations of the Stokes structure as $\sigma$ is varied. 

\section{Conclusions and Discussion}
\label{sect:CON}

In this work, we applied exponential asymptotic techniques to calculate asymptotic solutions to the discrete Airy equation \eqref{eqn:latAi}. The resultant asymptotic expressions contain both the higher-order Stokes phenomenon and Stokes curve accumulations; neither of these features are present in homogeneous linear second-order differential equations. In general, the higher-order Stokes phenomenon requires three interacting exponential contributions, while Stokes curve accumulations require a transseries with an infinite number of exponential terms. The process of discretization, even if it is linear, generates such a transseries solution and allows for the generation of effects that are restricted to higher-order or nonlinear continuous differential equations.

We showed that solutions to the discrete Airy equation \eqref{eqn:latAi} are described by the transseries \eqref{eqn:dtransser}. By comparing our solutions to the discrete Airy equation \eqref{eqn:latAi} with those of the continuous Airy equation \eqref{eqn:spAi}, we found that discretization significantly alters both the asymptotic form and the associated Stokes structure. Specifically, solutions to the discrete Airy equation \eqref{eqn:latAi} have an additional turning point at $x = -4/\sigma^2$ and an additional virtual turning point at $x = -2/\sigma^2$, which generate new Stokes curves that intersect at two distinct Stokes crossing points. These features are absent in the continuous case.

We also identified the higher-order Stokes phenomena associated with the Stokes crossing points. To the authors' knowledge, this is the first observation of higher-order Stokes switching arising in a discrete equation. In continuous equations, the higher-order Stokes phenomenon emerges from the intersection of Stokes curves \cite{chapman2005,howls2004}. Prior studies of discrete equations \cite{kashani2016} did not observe the higher-order Stokes phenomenon, because they study a different difference equation in which their corresponding Stokes curves did not intersect.

Because the asymptotic solution \eqref{eqn:dtransser} is an infinite-parameter transseries, we observed accumulations of both an infinite number of Stokes and anti-Stokes curves. The accumulation of curves in the Stokes structure is often associated with nonlinear differential equations and is an expected consequence of nonlinearity as seen in \cite{chapman2007}. Our results show that similar behavior can arise even in linear discrete systems with a finite number of turning points. This is a direct consequence of discretization, which necessarily introduces an infinite number of exponential contributions. We therefore conjecture that the accumulation of Stokes and anti-Stokes curves can be a generic feature of discrete equations, independent of nonlinearity.

In Section \ref{sect:difsig}, we studied how the Stokes structure changes as the spatial step $h = \sigma\eps$ takes complex values. We showed that variations in $\Arg(\sigma)$ significantly alter the Stokes structure. Our results align with earlier studies \cite{cao2014,li2020,wang2003,wang2005,wong1992b}, in which different asymptotic expansions were derived based on the relative positions of the turning points. 

We computed numerical solutions to the discrete Airy equation \eqref{eqn:latAi} that satisfy the boundary condition $y \to 0$ as $|x| \to \infty$, and compared them with our asymptotic solutions. We observed strong agreement, validating both the transseries solution \eqref{eqn:dtransser} and its associated Stokes structure. We also compared our findings with other asymptotic results for discrete equations in the literature, including those derived using series methods \cite{cao2014,li2020,wang2003,wang2005,wong1992b} and WKB methods \cite{burr2000,fedotov2019, kashani2016}. Our findings are consistent with these approaches in terms of turning point locations and the local asymptotic behavior near these points.

\section{Acknowledgements}

The authors would like to thank the Isaac Newton Institute for Mathematical Sciences, Cambridge, for support and hospitality during the programme ‘Applicable resurgent asymptotics: towards a universal theory’, where the work on this paper was conceptualised. The programme was supported by the EPSRC grant no. EP/R014604/1. AJMD and CJL gratefully acknowledge funding from ARC Discovery Project DP190101190. CJL gratefully acknowledges funding from ARC Discovery Project DP240101666.

\section{Roles}

\textbf{Aaron J. Moston-Duggan}: Conceptualisation, Formal Analysis, Visualisation, Writing - original draft. \textbf{Christopher J. Howls}: Writing - original draft, Writing - review \& editing. \textbf{Christopher J. Lustri}: Conceptualisation, Supervision, Writing - review \& editing.

\bibliographystyle{plain}
\bibliography{discAi.bib}

\appendix

\section{Factorial-Over-Power Analysis}
\label{sect:EA}

In this section, we show that the asymptotic solution \eqref{eqn:dtransser} to the discrete Airy equation \eqref{eqn:latAi}, previously derived by steepest descent, can also be obtained using exponential asymptotics based on factorial-over-power methods \cite{chapman1998, olde1995}. Unlike steepest descent, this approach does not require an integral representation and applies directly to the advance-delay equation \eqref{eqn:soadvAi}. Demonstrating this method is valuable, as it extends naturally to nonlinear problems and provides a useful tool for studying higher-order Stokes switching in nonlinear difference equations.

To apply the factorial-over-power methods, we expand the solution to \eqref{eqn:soadvAi} as a Taylor series about $\epsilon=0$ and obtain 
\begin{equation}
\label{eqn:2aitse}
    \frac{2}{\sigma^2}\sum_{j=1}^{\infty} \frac{\sigma^{2j}\epsilon^{2j}}{(2j)!}\diff{^{2j}y}{x^{2j}} - xy = 0 \, ,
\end{equation}
where $y(x)$ is now a local expression. We study equation \eqref{eqn:2aitse} using exponential asymptotic methods developed in \cite{chapman1998} and first applied to discrete systems in \cite{king2001}. 

\subsection{Methodology}
\label{sec:eamethod}

The steepest descent method is effective for deriving asymptotic solutions from integrals, however, many problems lack a convenient integral form. In such cases, exponential asymptotic methods \cite{chapman1998,olde1995} can be applied directly to the differential equation to determine the same asymptotic results as the steepest descent analysis.

We first expand the solution as an asymptotic power series of the form
\begin{equation}
    \label{eqn:asser}
        y \sim \left(\sum_{k=0}^\infty \epsilon^k A_k(x)\right)\mathrm{e}^{-S(x)/\epsilon} \quad\quad \text{as} \quad\quad \epsilon \to 0 \, .
\end{equation}
We substitute the asymptotic series \eqref{eqn:asser} into the governing equation and match the terms for all orders of $\epsilon$. Solving these equations for each order of $\epsilon$ gives the values of $S$ and $A_k$. 

Determining $A_k$ requires repeated differentiation of earlier terms. When these terms are singular, this causes ``factorial-over-power'' divergence \cite{dingle1973}, enabling an asymptotic description of the late-order terms as $k \to \infty$. Based on \cite{dingle1973}, the authors of \cite{chapman1998} proposed an ansatz for late-order terms, the leading-order term is a sum of terms of the form
\begin{equation}
    \label{eqn:LOT}
        A_k \sim \dfrac{B(x)\Gamma(k + \gamma)}{\chi(x)^{k + \gamma}} \quad\quad \text{as}\quad\quad k\to\infty \, ,
\end{equation}
where $\Gamma$ is the gamma function, $\gamma$ is a constant and $B$ and $\chi$ are functions of $x$. Each term in the sum is associated with a particular singularity of $A_0$. The contribution for each singularity can be determined independently, taking the sum of these contributions provides the complete behavior of the late-order terms \cite{dingle1973}.

We call the functions $B$ the prefactor and $\chi$ the singulant. The singulants satisfy $\chi = 0$ at the singularities of $A_0$, ensuring that $A_k$ is also singular at the same location. Substituting the ansatz \eqref{eqn:LOT} into the recurrence relation and matching terms as $k \to \infty$ determines $B$ and $\chi$. The value of $\gamma$ is chosen to ensure consistency of late-order terms \eqref{eqn:LOT} with the leading-order solution $A_0$ near the singularities.

The singulant $\chi$ determines the location of Stokes curves. As shown by \cite{dingle1973}, a Stokes curve associated with a change in exponentially small behaviour that is switched by a power series expansion satisfies the conditions
\begin{equation}
    \label{eqn:eastokcon}
        \Im(\chi) = 0 \quad \text{and} \quad  \Re(\chi)>0 \, .
\end{equation}
The first condition ensures the dominant and subdominant exponential contributions have equal phase, while the second restricts Stokes switching to exponentially small contributions.

Knowing the late-order form \eqref{eqn:LOT} allows optimal truncation of the divergent series \eqref{eqn:asser} at the smallest term, typically producing an exponentially small error as $\epsilon \to 0$ \cite{berry1988, berry1989, boyd1999}. The optimally truncated series is given as
\begin{equation}
    \label{eqn:optAi}
        y = \left(\sum_{k=0}^{K-1} \epsilon^k A_k + R_{K} \right)         \mathrm{e}^{-S/\epsilon}  \quad \text{as} \quad \epsilon\to0 \, ,
\end{equation}
where the remainder $R_{K}$ is exponentially small as $\epsilon\to 0$. The optimal truncation point $K$ occurs at the term of least magnitude \cite{boyd1999}. Applying this heuristic to \eqref{eqn:LOT} gives $K \sim |\chi|/\epsilon$.

We determine the exponentially small remainder $R_{K}$ by substituting \eqref{eqn:optAi} into the original equation. Away from the Stokes curves, we approximate $R_{K}$ using the WKB ansatz \cite{bender2013}. The WKB approximation breaks down near the Stokes curve. Near the Stokes curve, we determine $R_{K}$ using the variation of parameters ansatz 
\begin{equation}
\label{eqn:remter}
        R_{K}\sim \mathcal{M}(x)B(x)\mathrm{e}^{-\chi(x)/\epsilon} \quad \text{as} \quad \epsilon \to 0 \, ,
\end{equation}
where $\mathcal{M}(x)$, is called the Stokes multiplier, which varies rapidly in a width $\mathcal{O}(\epsilon^{1/2})$ around the Stokes curve and encodes the Stokes switching. We compute $\mathcal{M}(x)$ using the matched asymptotic expansion procedure from \cite{olde1995}.

\subsection{Series expansion}

Substituting the ansatz \eqref{eqn:asser} into \eqref{eqn:2aitse} to obtain
\begin{equation}
\label{eqn:2aitse2}
    \frac{2}{\sigma^2}\sum_{j=1}^{\infty}\sum_{k=0}^{\infty} \frac{\sigma^{2j}\epsilon^{2j+k}}{(2j)!}\diff{^{2j}}{x^{2j}}\left(A_k \e^{-S/\eps}\right) - x\sum_{k=0}^{\infty} \eps^k A_k \e^{-S/\eps}= 0 \, .
\end{equation}
We apply the general Leibniz rule and Fa\'{a} di Bruno's formula to write
\begin{equation}
\label{ident1}
\diff{^{2j}}{x^{2j}}\left(A_k \mathrm{e}^{-S/\epsilon}\right) = \sum_{l=0}^{2j}\frac{(2j)!}{l!(2j-l)!}\diff{^{2j-l}A_k}{x^{2j-l}} \sum_{m=0}^{l} \frac{(-1)^m \mathcal{B}_{l,m}^S}{\epsilon^m} \e^{-S/\eps} \, ,
\end{equation}
where $\mathcal{B}^S_{l,m}$ are the partial Bell polynomials \cite{roman1978} where the superscript notation indicates the argument, such that
\begin{equation}
    \mathcal{B}^S_{l,m} = \mathcal{B}_{l,m}\left(\diff{S}{x},\diff{^2S}{x^2},\cdots,\diff{^{l-m+1}S}{x^{l-m+1}}\right) \, .
\end{equation}
The partial Bell polynomials allow our expressions to be written compactly. We will make use of the identities
\begin{equation}\label{eq:bellidentities}
\mathcal{B}^S_{l,l} = \left(\diff{S}{x}\right)^l \quad \text{and} \quad \mathcal{B}^S_{l,l-1} = \binom{l}{2}\left(\diff{S}{x}\right)^{l-2}\diff{^2S}{x^2} \, .
\end{equation}
We apply \eqref{ident1} to \eqref{eqn:2aitse2} to obtain
\begin{equation}
\label{eqn:fwkbsoadvex}
    \frac{2}{\sigma^2} \sum_{j=1}^\infty \sum_{k=0}^\infty \sum_{l=0}^{2j} \sum_{m=0}^l \frac{(-1)^m\sigma^{2j}\epsilon^{2j+k-m}\mathcal{B}_{l,m}^S }{l!(2j-l)!}  \frac{\mathrm{d}^{2j-l}A_{k}}{\mathrm{d}x^{2j-l}} - x\sum_{k=0}^\infty \epsilon^k A_{k} = 0 \, .
\end{equation}

\subsubsection{Exponent equation} 

We balance terms in \eqref{eqn:fwkbsoadvex} at $\mathcal{O}(1)$ as $\epsilon\to0$. After using the identities from \eqref{eq:bellidentities}, we obtain  
\begin{align}
0 = \frac{2}{\sigma^2}\sum_{j=1}^\infty \frac{\sigma^{2j}}{(2j)!}\left(\frac{\mathrm{d}S}{\mathrm{d}x}\right)^{2j} - x = \frac{2}{\sigma^2}\left(\cosh\left(\sigma\diff{S}{x}\right)-1\right) - x \,.
\label{eqn:expfacs}
\end{align}
Solving this equation gives two families of solution for $S$ 
\begin{equation}
\label{serxexp1}
    S_{s}^{\pm} = -\frac{\mathrm{i}}{\sigma}\left[\left(x+\frac{2}{\sigma^2}\right)\left(\pm\mathrm{i}\cosh^{-1}\left(1+\frac{\sigma^2 x}{2}\right)  -2\pi s\right) \mp\i\sqrt{x\left(\sigma^2 x+ 4\right)}\right].
\end{equation}
We label the families by a superscript for the sign choice and a subscript for the index $s$. The exponents $S_s^\pm$ \eqref{serxexp1} correspond to the saddle heights $\phi_s^\pm$ \eqref{e:sheights} from the steepest descents analysis. The solution then takes the form
\begin{equation}\label{e:transseries0}
    y \sim \sum_{s=-\infty}^{\infty}\left[c_s^+ \sum_{k=0}^{\infty} \epsilon^k A_{s,k}^+ \mathrm{e}^{S_s^+/\epsilon} + c_s^- \sum_{k=0}^{\infty} \epsilon^k A_{s,k}^- \mathrm{e}^{S_s^-/\epsilon}  \right] \quad \mathrm{as} \quad \eps \to 0 \, ,
\end{equation}
where $c_{s}^\pm$ are constants to be specified in one of the regions $\mathcal{D}_1$, $\mathcal{D}_2$, or $\mathcal{D}_3$ to obtain particular solutions to \eqref{eqn:soadvAi}.

\subsubsection{Leading-order solution}

We continue to balance terms in equation \eqref{eqn:fwkbsoadvex} of size $\mathcal{O}(\epsilon)$ as $\epsilon\to 0$ in order to determine the form of the leading-order terms in \eqref{e:transseries0}, $A_{s,0}^{\pm}$. Simplifying the resultant expression using \eqref{eq:bellidentities} and \eqref{eqn:expfacs} gives
\begin{align}
  0 & = \sum_{j=1}^\infty \frac{\sigma^{2j}}{(2j-1)!}\left(\frac{\mathrm{d}S_s^{\pm}}{\mathrm{d}x}\right)^{2j-1}\frac{\mathrm{d}A_{s,0}^{\pm}}{\mathrm{d}x} +   \frac{\mathrm{d}^2S_s^{\pm}}{\mathrm{d}x^2}\sum_{j=1}^\infty \frac{\sigma^{2j}}{2!(2j-2)!}\left(\frac{\mathrm{d}S_s^{\pm}}{\mathrm{d}x}\right)^{2j-2}A_{s,0}^{\pm}  \\
  & = \sinh\left(\sigma\diff{S_s^{\pm}}{x}\right)\diff{A_{s,0}^{\pm}}{x} + \frac{\sigma}{2}\cosh\left(\sigma\diff{S_s^{\pm}}{x}\right)\diff{^2 S_s^{\pm}}{x^2}A_{s,0}^{\pm} \, . \label{2aipref0}
  \end{align}
We solve equation \eqref{2aipref0} to obtain 
\begin{equation}
\label{dprefo}
    A_{s,0}^{\pm} = \frac{1}{\sqrt{2\pi\eps}x^{1/4}(\sigma^2x+4)^{1/4}} \, ,
\end{equation}
where the multiplicative constants are included for later algebraic convenience. 

The leading-order behaviour of each exponential contribution 
as $\epsilon \to 0$ is therefore given by
\begin{align}
\label{dleado1}
    y_{s,0}^{\pm} & =  \frac{1}{\sqrt{2\pi\epsilon}x^{1/4}(\sigma^2x+4)^{1/4}} \mathrm{e}^{\left(\frac{\mathrm{i}}{\sigma \epsilon}\left[\left(x+\frac{2}{\sigma^2}\right)\left(\pm\mathrm{i}\cosh^{-1}\left(1+\frac{\sigma^2x}{2}\right)  + 2\pi s \right) \mp \mathrm{i}\sqrt{x\left({\sigma^2}x+4\right)}\right]   \right)} \, .
\end{align}
Each contribution has two singularities, giving the turning points $x=-4/\sigma^2$ and $x=0$, in agreement with those identified from the steepest descent analysis. 

\subsection{Late-order terms} \label{s:LOT}

To calculate all series terms $A^{\pm}_{s,k}$, we match the terms in \eqref{eqn:fwkbsoadvex} at each power of $\eps$ and solve each equation recursively. Since the analysis applies identically to all $S_s^{\pm}$, we simplify the notation by writing $S_s^{\pm}=S$ and $A^\pm_{s,k} = A_k$.

Balancing terms in \eqref{eqn:fwkbsoadvex} at order $\mathcal{O}(\epsilon^q)$ with $q \geq 2$ as $\epsilon \to 0$ gives
\begin{align}
\label{drecurra2}
\frac{2}{\sigma^2}&\sum_{j=\lfloor \frac{q-1}{2} \rfloor +1}^{\infty}  \sum_{m=2j-q}^{2j}  \sum_{l=m}^{2j}  \frac{(-1)^{m}\sigma^{2j}\mathcal{B}_{l,m}^S}{l!(2j-l)!}\diff{^{2j-l}A_{m+q-2j}}{x^{2j-l}} \nonumber \\  & \hspace{2.8cm} +  \frac{2}{\sigma^2}\sum_{j=1}^{\lfloor \frac{q-1}{2} \rfloor}\sum_{m=0}^{2j}\sum_{l=m}^{2j} \frac{(-1)^{m}\sigma^{2j}\mathcal{B}_{l,m}^S}{l!(2j-l)!}\diff{^{2j-l}A_{m+q-2j}}{x^{2j-l}}  - xA_q = 0 \, ,
\end{align}
where $\lfloor \cdot \rfloor$ denotes the floor function. Each contribution $y_{s,0}^\pm$ \eqref{dleado1} is singular at 
$x = -4/\sigma^2$ and $x = 0$, so the asymptotic series \eqref{eqn:asser} exhibit factorial-over-power divergence. Hence, the terms $A_{k}$ are described by the late-order ansatz \eqref{eqn:LOT}. We apply \eqref{eqn:LOT} to the recurrence relation \eqref{drecurra2} to obtain
\begin{align} \nonumber
 \frac{2}{\sigma^2}  \sum_{j=\lfloor \frac{q-1}{2} \rfloor + 1}^{\infty}\sum_{m=2j-q}^{2j}\sum_{l=m}^{2j}\sum_{n=0}^{2j-l}\sum_{p=0}^{n} \frac{\Gamma(m+q+p+\gamma-2j)}{\chi^{m+q+p+\gamma-2j}}  \frac{(-1)^{m+p}\sigma^{2j}\mathcal{B}_{l,m}^S \mathcal{B}_{n,p}^\chi }{l!n!(2j-l-n)!} \diff{^{2j-l-n}B}{x^{2j-l-n}}&  \\  + \frac{2}{\sigma^2}  \sum_{j=1}^{\lfloor \frac{q-1}{2} \rfloor}\sum_{m=0}^{2j}\sum_{l=m}^{2j}\sum_{n=0}^{2j-l}\sum_{p=0}^{n} \frac{\Gamma(m+q+p+\gamma-2j)}{\chi^{m+q+p+\gamma-2j}}  \frac{(-1)^{m+p}\sigma^{2j}\mathcal{B}_{l,m}^S \mathcal{B}_{n,p}^\chi}{l!n!(2j-l-n)!} \diff{^{2j-l-n}B}{x^{2j-l-n}}& \nonumber \\
  -xB\frac{\Gamma(q+\gamma)}{\chi^{q+\gamma}} &= 0 \, . \label{drecurralot}
 \end{align}
 As $q \to \infty$, the first term will not contribute in any subsequent analysis and is henceforth neglected.

\subsubsection{Singulant equation}

We balance the largest terms in equation \eqref{drecurralot}, which are of order $\mathcal{O}(A_q)$ as $q\to \infty$ to obtain
\begin{align}
\label{eqn:dsinglab}
\frac{2}{\sigma^2}  \sum_{j=1}^{\lfloor \frac{q-1}{2} \rfloor}\sum_{m=0}^{2j} & \frac{\sigma^{2j}}{m!(2j-m)!}\left(\diff{S}{x}\right)^m\left(\diff{\chi}{x}\right)^{2j-m} - x = 0 \, .
\end{align}
We are considering the behavior of late-order terms \eqref{eqn:LOT}, and therefore the large-$q$ limit of \eqref{eqn:dsinglab}. Extending the sums to infinity yields the leading-order behavior of \eqref{eqn:dsinglab} as $q\to\infty$, which gives
\begin{align}
\label{eqn:dsinglab2}
   0 &=  \frac{2}{\sigma^2}\sum_{j=1}^\infty \sum_{m=0}^{2j} \frac{\sigma^{2j}}{m!(2j-m)!} \left( \diff{S}{x}\right)^{m}  \left(\diff{\chi}{x}\right)^{2j-m} - x \\ 
   &= \frac{2}{\sigma^2} \left( \cosh \left(\sigma\left(\diff{S}{x}+\diff{\chi}{x} \right)\right) -1 \right)  - x \, .
\end{align}
Solving \eqref{eqn:dsinglab2} for $\mathrm{d}\chi/\mathrm{d}x$ we find two singulants for each exponent $S_{s}^-$ and $S_{s}^+$. We denote these singulants as
\begin{align}
\label{dchi2}
    \diff{\chi_{s,p}^{\pm,\pm}}{x} = \frac{2\pi\mathrm{i}(s-p)}{\sigma} \, , \qquad \diff{\chi_{s,p}^{\pm,\mp}}{x} = \frac{2}{\sigma}\left(\pi\mathrm{i}(s-p) \pm \cosh^{-1}\left(1+\frac{\sigma^2x}{2}\right) \right), 
\end{align}
where upper and lower sign choices correspond. The singulants $\chi^{-,-}_{s,p}$, $\chi^{-,+}_{s,p}$, $\chi^{+,-}_{s,p}$,and $\chi^{+,+}_{s,p}$ correspond to the Stokes switching behaviour across the Stokes curves $\mathcal{S}^{-,-}_{s,p}$, $\mathcal{S}^{-,+}_{s,p}$, $\mathcal{S}^{+,-}_{s,p}$, and $\mathcal{S}^{+,+}_{s,p}$ respectively. The first superscript and subscript indicates the dominant exponential contribution, while the second indicates the exponential term that is switched on. For example, the singulant $\chi^{+,-}_{s,p}$ is associated with the $y_s^+$ contribution switching on the $y_p^-$ contribution.

Although the series analysis can identify active Stokes curves, care is needed due to the many possible interacting contributions. To streamline the analysis, we instead use the steepest descent results to focus only on those Stokes curves known to be active in the asymptotic solution.

The Stokes curves $\mathcal{S}_{s,s}^{-,+}$ originate at the turning point $x = 0$. Solving \eqref{dchi2} with the condition $\chi = 0$ at $x = 0$ shows that the corresponding singulants are 
\begin{equation}\label{e:singss}
    \chi_{s,s}^{+,-} =\frac{2\mathrm{i}}{\sigma}\left[ \left(x+\frac{2}{\sigma^2}\right)\mathrm{i}\cosh^{-1}\left(1+\frac{\sigma^2 x}{2}\right) - \mathrm{i}\sqrt{x\left({\sigma^2}x+4\right)} \, \right] \, .
\end{equation} 
The Stokes curves $\mathcal{S}_{s,s+1}^{-,+}$ originate at the turning point $x = -4/\sigma^2$. Solving \eqref{dchi2} with the condition $\chi = 0$ at $x = -4/\sigma^2$ shows that the corresponding singulants are 
\begin{equation}
    \chi_{s,s+1}^{-,+} = -\frac{2\mathrm{i}}{\sigma}\left[ \left(x+\frac{2}{\sigma^2}\right)\left(\mathrm{i}\cosh^{-1}\left(1+\frac{\sigma^2 x}{2}\right) + 2\pi \right) -\mathrm{i}\sqrt{x\left({\sigma^2}x+4\right)} \, \right] \, .
\end{equation}
The Stokes curves $\mathcal{S}_{s,s+1}^{-,-}$ and $\mathcal{S}_{s,s+1}^{+,+}$ originate at the virtual turning point $x = -2/\sigma$. Solving \eqref{dchi2} with $\chi = 0$ at $x = -2/\sigma^2$ shows that the corresponding singulants are 
\begin{equation}
    \chi_{s,s+1}^{-,-} = \chi_{s,s+1}^{+,+} = \frac{2\pi\mathrm{i}}{\sigma}\left(x+\frac{2}{\sigma^2}\right) \, .
\end{equation}

Each singulant can be directly related to the corresponding steepest descent expression by noting that $\chi^{+,-}_{s,p} = \phi^{+}_s - \phi^{-}_p$, with the other singulants obtained by adjusting signs and indices accordingly. Applying condition \eqref{eqn:eastokcon} to each singulant yields the same Stokes curves as those found using the steepest descent method in Section \ref{sect:SD}.

\subsubsection{Prefactor equation}

We balance the next to largest terms in equation \eqref{drecurralot}, which are of order $\mathcal{O}(A_{q-1})$ as $q\to \infty$ to obtain
\begin{align}
\label{dpreflab}
\frac{2}{\sigma^2}  \sum_{j=1}^{\lfloor \frac{q-1}{2} \rfloor}&\sum_{m=0}^{2j-1}\sum_{l=m}^{m+1}\sum_{n=2j-m-1}^{2j-l} \frac{\sigma^{2j}\mathcal{B}_{l,m}^S \mathcal{B}_{n,2j-m-1}^\chi }{l!n!(2j-l-n)!} \diff{^{2j-l-n}B}{x^{2j-l-n}}= 0 \, .
\end{align}
We take the limit $q\to \infty$ in equation \eqref{dpreflab} and apply the identities from \eqref{eq:bellidentities}, to obtain
\begin{align}
\label{dpreflab23}
\sinh\left(\sigma\left(\diff{S}{x}+\diff{\chi}{x}\right)\right)\diff{B}{x} + \frac{\sigma}{2}\left(\diff{^2S}{x^2}+\diff{^2\chi}{x^2} \right)\cosh\left(\sigma\left(\diff{S}{x}+\diff{\chi}{x} \right)\right)B = 0 \, .
\end{align}
We solve the prefactor equation \eqref{dpreflab23} for each of the singulants, giving
\begin{align}
\label{dpref}
B_s^\pm = \frac{ C^\pm_s }{\sqrt{2\pi\epsilon}x^{1/4}(\sigma^2x+4)^{1/4}} \, .
\end{align}
where $C^{\pm}_s$ are constants that remain to be determined. The multiplicative constants in \eqref{dpref} are chosen for algebraic convenience, such that $C_s^{\pm} = 1$ in the final asymptotic expression.

\paragraph{Calculating \texorpdfstring{$C^+_s$}{C+s}, \texorpdfstring{$C^-_s$}{C-s} and \texorpdfstring{$\gamma$}{gamma}:}

To determine $C_s^+$, we match the late-order ansatz \eqref{eqn:LOT} with a local expansion of the solution near the singularity $x=0$. This is necessary because the outer expansion \eqref{eqn:asser} of the late-order terms \eqref{eqn:LOT} breaks down in the region where $\epsilon^{k}A_k \mathrm{e}^{-S/\eps}  \sim \epsilon^{k+1}A_{k+1} \mathrm{e}^{-S/\eps}$; that is, where $\chi = \mathcal{O}(\epsilon)$ as $k\to\infty$. 

The singulant and the leading-order contribution satisfy
\begin{equation}
    \chi^{+,-}_{s,s} \sim \frac{4x^{3/2}}{3} \quad\quad \text{and} \quad\quad A_0 \mathrm{e}^{-S^+_s} \sim \frac{C^+_s}{2\sqrt{\pi\epsilon}x^{1/4}} \mathrm{e}^{\frac{2\mathrm{i}\pi}{\sigma}\left(x+\frac{2}{\sigma^2}\right) - \frac{2}{3}x^{3/2}} \quad\quad \text{as} \quad\quad x\to 0 \, .
\end{equation}
The envelope for the leading-order contribution and the singulant match those of the singularly-perturbed Airy equation \eqref{eqn:spAi}. Furthermore, applying the inner scaling $x = \epsilon^{2/3}\eta$ and  $\phi(\eta)=\epsilon^{-1/6}y(x)$, shows that the local equation for the discrete Airy equation \eqref{eqn:2aitse} is the same as that of the continuous Airy equation \eqref{eqn:spAi}. 

The analysis therefore proceeds identically to that of the well-known Airy equation \eqref{eqn:spAi}. For brevity, we omit repeating this analysis, and use the result that $C^+_s = 1$. A similar analysis shows $C^-_s = 1$.

The strength of the singularity in the late-order terms \eqref{eqn:LOT} must be consistent with the leading-order solutions \eqref{dleado1} near the singularities $x=0$ and $x=-4/\sigma^2$. The leading-order solution \eqref{dleado1} has singularities of order 1/4 at these points. Since $B$ is also singular with order 1/4, the late-order terms \eqref{eqn:LOT} are consistent with the leading-order solutions \eqref{dleado1} if and only if $\gamma=0$.

\subsection{Stokes switching}

We apply the exponential asymptotic method developed in \cite{olde1995}, optimally truncating the divergent series \eqref{eqn:asser} after $K$ terms, yields the expression \eqref{eqn:optAi}, where $R_K$ denotes the remainder. Following the heuristic of \cite{boyd1999, dingle1973}, the optimal truncation point is the value of $K$ for which consecutive terms in the series have the same magnitude in the limit $\epsilon \to 0$. This typically occurs after an asymptotically large number of terms, so we use the late-order ansatz \eqref{eqn:LOT} to estimate $K$. The optimal truncation point satisfies $K \sim |\chi|/\epsilon$ as $\epsilon \to 0$, which justifies the use of the late-order ansatz. We therefore set $K = |\chi|/\epsilon + \omega$, where $\omega \in [0,1)$ is chosen to ensure that $K$ is an integer.

This follows directly from the method in \cite{olde1995}; however, it is not obvious that the governing equation \eqref{eqn:soadvAi} reduces to the standard form seen in exponential asymptotics. We therefore outline the the details here. The analysis is presented for general $\chi$ and $B$, so it applies uniformly to each switching contribution in the solution.

We substitute the optimally truncated series \eqref{eqn:optAi} into \eqref{eqn:soadvAi} to obtain
\begin{align}
\label{eqn:opttrunc}
    \frac{2}{\sigma^2} \sum_{j=1}^\infty \sum_{l=0}^{2j} \sum_{m=0}^l & \frac{(-1)^m\sigma^{2j}\epsilon^{2j-m}\mathcal{B}_{l,m}^S}{l!(2j-l)!} \diff{^{2j-l}R}{x^{2j-l}} - x R = \nonumber \\ &  -\left(  \displaystyle{\frac{2}{\sigma^2} \sum_{j=1}^\infty \sum_{k=0}^{K-1} \sum_{l=0}^{2j} \sum_{m=0}^l \frac{(-1)^m\sigma^{2j}\epsilon^{2j+k-m}\mathcal{B}_{l,m}^S}{l!(2j-l)!} \frac{\mathrm{d}^{2j-l}A_{k}}{\mathrm{d}x^{2j-l}}  - x\sum_{k=0}^{K-1} \epsilon^k A_{k}} \right) \, .
\end{align}
Simplifying \eqref{eqn:opttrunc} using \eqref{eqn:expfacs}, \eqref{2aipref0}, and the recurrence relation \eqref{drecurra2}, we obtain the leading-order terms, we obtain as $\eps \to 0$ that
\begin{align}
\label{eqn:opttrunc2}
        \frac{2}{\sigma^2} \sum_{j=1}^\infty \sum_{l=0}^{2j} \sum_{m=0}^l  \frac{(-1)^m\sigma^{2j}\epsilon^{2j-m}\mathcal{B}_{l,m}^S}{l!(2j-l)!}  \diff{^{2j-l}R}{x^{2j-l}} - x R \sim -\frac{2\epsilon^{K+1}}{\sigma^2}\sum_{j=1}^\infty \sum_{l=2j-1}^{2j} \frac{\sigma^{2j}\mathcal{B}_{l,2j-1}^S}{l!(2j-l)!}\diff{^{2j-l}A_{K}}{x^{2j-l}}\,,
\end{align}
where the terms omitted in \eqref{eqn:opttrunc2} are at most $\mathcal{O}(\epsilon^{K+2})$ as $\epsilon \to 0$ and are therefore negligible. 

Since $K \to \infty$ as $\epsilon \to 0$, we insert the late-order ansatz \eqref{eqn:LOT} into \eqref{eqn:opttrunc2} and retaining only the largest terms as $\epsilon\to0$, we obtain in this limit
\begin{align}
\label{eqn:opttrunc3}
    \frac{2}{\sigma^2} \sum_{j=1}^\infty \sum_{l=0}^{2j} \sum_{m=0}^l  \frac{(-1)^m\sigma^{2j}\epsilon^{2j-m}\mathcal{B}_{l,m}^S}{l!(2j-l)!}  \diff{^{2j-l}R}{x^{2j-l}} - xR   \sim \frac{2\epsilon^{K+1}}{\sigma}\sinh\left(\sigma\diff{S}{x}\right)\diff{\chi}{x}\frac{B\Gamma(K+1)}{\chi^{K+1}}\, ,
\end{align}
where terms omitted from \eqref{eqn:opttrunc3} have a size of at most $\mathcal{O}(\Gamma(K)/\chi^{K})$ as $K \to \infty$ and are negligible in this analysis.

The right-hand side of \eqref{eqn:opttrunc3} is small compared to the terms on the left-hand side, except within a region of width $\mathcal{O}(\sqrt{\epsilon})$ around the curve $\Im(\chi) = 0$, which is the Stokes curve. Away from the Stokes curve, we use a WKB ansatz to solve the homogeneous version of \eqref{eqn:opttrunc3}, given by
\begin{equation}
\label{eqn:opttrunchom}
    \frac{2}{\sigma^2} \sum_{j=1}^\infty \sum_{l=0}^{2j} \sum_{m=0}^l\frac{(-1)^m\sigma^{2j}\epsilon^{2j-m}\mathcal{B}_{l,m}^S}{l!(2j-l)!}\diff{^{2j-l}R_{K}}{x^{2j-l}} - x R_{K}  = 0 \, .
\end{equation}

The WKB analysis of \eqref{eqn:opttrunchom} motivates a variation of parameters approach using the ansatz \eqref{eqn:remter} to describe the behavior of $R_K$ near the Stokes curve. The Stokes multiplier $\mathcal{M}$ varies locally around the Stokes curve, where the right-hand side of \eqref{eqn:opttrunc3} is not negligible, capturing the switching behavior. Substituting the ansatz \eqref{eqn:remter} into \eqref{eqn:opttrunc3}, and simplifying using \eqref{eqn:expfacs}, \eqref{eqn:dsinglab2}, and \eqref{dpref}, gives the Stokes multiplier equation
\begin{equation}
\label{eqn:stokmult2}
    \diff{ \mathcal{M}}{x} \sim  -\frac{ \sinh \left( \sigma \diff{S}{x} \right) \diff{\chi}{x}}{ \sinh \left( \sigma\left( \diff{S}{x} + \diff{\chi}{x} \right)\right) } \frac{ \epsilon^{K} \Gamma(K+1)}{ \chi^{K+1} }\mathrm{e}^{\chi/\epsilon} \quad \text{as} \quad \epsilon \to 0 \, .
\end{equation}
The right-hand side of \eqref{eqn:stokmult2} simplifies since the first term is either $1$ or $-1$, depending on $\chi$, hence, we write
\begin{equation}
\label{eqn:stokmult3}
    \diff{\mathcal{M}}{x} \sim  \mp \diff{\chi}{x}\frac{ \epsilon^{K} \Gamma(K+1)}{ \chi^{K+1} }\mathrm{e}^{\chi/\epsilon} \quad \text{as} \quad \epsilon \to 0 \, .
\end{equation}
It is useful to make $\chi$ the independent variable. Applying this change of variables to \eqref{eqn:stokmult2} gives
\begin{equation}
\label{eqn:stokmult3a}
    \diff{\mathcal{M}}{\chi} \sim \mp \frac{ \epsilon^{K} \Gamma(K+1)}{ \chi^{K+1} }\mathrm{e}^{\chi/\epsilon} \quad \text{as} \quad \epsilon \to 0 \, .
\end{equation}
From this point, the analysis follows the standard matched asymptotic expansion method from \cite{olde1995}, which gives
\begin{equation}
\label{eqn:dairy:stokmul9}
    \mathcal{M} \sim \pm \mathrm{i}\pi \erf\left(\sqrt{\frac{|\chi|}{\epsilon}}\Arg(\chi)\right) + C \quad \text{as} \quad \epsilon \to 0 \, ,
\end{equation}
where “$\Arg$” denotes the principal argument. This is the standard form of Stokes switching encountered in steepest descents analysis, confirming agreement with the results of Section \ref{sect:SD}. Hence, the asymptotic behavior of the solution to the discrete Airy equation \eqref{eqn:latAi} can be determined using only asymptotic series methods.

\end{document}